\documentclass[apj,iop]{emulateapj}
\usepackage{times,mathptmx}

\shorttitle{{\sc Binary satellite migrating in a gaseous disk}}
\shortauthors{{\sc Baruteau, Cuadra \& Lin}}

\begin{document}
\title{Binaries migrating in a gaseous disk: Where are the Galactic
  center binaries?}  \author{C. Baruteau\altaffilmark{1},
  J. Cuadra\altaffilmark{2,3,4} and D.N.C. Lin\altaffilmark{1,5}}
\affil{$^1$Department of Astronomy and Astrophysics, University of California, Santa Cruz, CA 95064, USA\\
  $^2$Shanghai Astronomical Observatory, Shanghai 200030, China\\
  $^3$Max-Planck-Institut f\"ur Astrophysik, D-85741 Garching, Germany\\
  $^4$Departamento de Astronom\'ia y Astrof\'isica, Pontificia Universidad Cat\'olica de Chile, Santiago, Chile\\
  $^5$Kavli Institute of Astronomy and Astrophysics, Peking
  University, Beijing, China } \email{clement.baruteau@ucolick.org;
  jcuadra@astro.puc.cl; lin@ucolick.org}


\begin{abstract}
  The massive stars in the Galactic center inner arcsecond share
  analogous properties with the so-called Hot Jupiters. Most of these
  young stars have highly eccentric orbits, and were probably not
  formed in-situ. It has been proposed that these stars acquired their
  current orbits from the tidal disruption of compact massive binaries
  scattered toward the proximity of the central supermassive black
  hole. Assuming a binary star formed in a thin gaseous disk beyond
  $0.1$ pc from the central object, we investigate the relevance of
  disk--satellite interactions to harden the binding energy of the
  binary, and to drive its inward migration.  A massive, equal-mass
  binary star is found to become more tightly wound as it migrates
  inwards toward the central black hole. The migration timescale is
  very similar to that of a single-star satellite of the same
  mass. The binary's hardening is caused by the formation of spiral
  tails lagging the stars inside the binary's Hill radius. We show
  that the hardening timescale is mostly determined by the mass of gas
  inside the binary's Hill radius, and that it is much shorter than
  the migration timescale. We discuss some implications of the
  binary's hardening process. When the more massive (primary)
  components of close binaries eject most their mass through supernova
  explosion, their secondary stars may attain a range of
  eccentricities and inclinations.  Such processes may provide an
  alternative unified scenario for the origin of the kinematic
  properties of the central cluster and S-stars in the Galactic center
  as well as the high velocity stars in the Galactic halo.
\end{abstract}

% ========================
\section{Introduction}
% ========================
The supermassive black hole (SMBH) in the Galactic center is
surrounded by a parsec-scale cluster of young and massive stars. At
separations larger than $\sim 1''$ or $0.04$ pc \footnote{Given the
  heliocentric distance to the Galactic center of $\sim 8$ kpc.},
about 100 OB stars are observed in a moderately thin disk that rotates
clockwise on the sky and extends up to $\sim 0.6$ pc. The disk's scale
height to radius ratio is $\sim 0.1$. The stars have masses $\gtrsim
10\;M_{\odot}$ and a common age $\sim 6$ Myr. Their eccentricities
range up to 0.8, with a typical value $\sim 0.4$ \citep[e.g.,][]{lu09,
  bartko09, bartko09l}. There is observational evidence for the
existence of a second, less massive counter-clockwise disk comprising
stars with similar age and kinematic properties as in the clockwise
disk \citep{paumard06, bartko09, bartko09l}. In contrast, stars within
the central arcsecond have orbits with random orientations and even
higher eccentricities. Known as the ``S-stars'', they are
main-sequence B stars with masses of $\sim 10\;M_{\odot}$ and a
lifespan of several $\times 10^{7}$ yrs \citep[e.g.,][]{schoedel02,
  ghez03, bartko09l}. The orbits of about $20$ of them have been
determined by \cite{gillessen09}. Their semi-major axes are as small
as $\sim 4\times 10^{-3}$ pc $\approx 800$ AU, and their
eccentricities are typically $\gtrsim 0.8$.

Explaining the presence of such young stars so close to the SMBH is
challenging. The standard star formation scenario, invoking the
collapse of self-gravitating cold molecular clouds, is unlikely to
occur, because of the strong tidal perturbation from the SMBH
\citep{morris93}. Nonetheless, the tidal disruption of a molecular
cloud can form a thin accretion disk around the central object
\citep{br08}. In-situ star formation may then be triggered when the
disk density is high enough so its self-gravity overcomes the tidal
force of the SMBH \citep{lb03, nc05}, a process first predicted for
the massive accretion disks present in AGNs \citep[e.g.,][]{ks80,
  sb89, goodman03}.

Alternatively, stars may have formed far from their current location
and have migrated inwards due to dynamical friction. While the
migration time for individual stars is much too long, a massive
($\gtrsim 10^5\;M_{\odot}$), dense stellar cluster formed several
parsecs away could reach the central parsec in a few $10^6$ yrs
\citep[e.g.,][]{gerhard01}.  Both the self-gravitating disk scenario
and the infalling cluster scenario can produce stellar disks
\citep[see][and references therein]{yll07}, but the observed
properties of the stellar disk in the central parsec favor in-situ
formation \citep{ns05, paumard06, lu09, bartko09}.

Additional complexity arises when attempting to reproduce the peculiar
orbits of the S-stars, which are even closer to the SMBH.  The most
promising mechanism involves the three-body interaction of a binary
star and the SMBH \citep{Hills88}. In particular, \cite{gq03} proposed
that the S-stars were initially in compact binary systems. The binary
can get tidally disrupted if it is scattered sufficiently close to the
SMBH, such that its binary separation is a significant fraction of its
Hill radius at its closest approach to the SMBH.  In addition to
potentially explaining the orbits of the S-stars, the binary's tidal
disruption scenario could account for the observed hypervelocity stars
\citep{Hills88, YT03, gualandris05, LuYuLin07, peretshv09}, most of
which are B-type stars, just like the S-stars
\citep[e.g.,][]{Brownetal07}. In this context, \cite{lzy10a} and
\cite{lzy10b} have recently pointed out that the spatial distribution
of the hypervelocity stars is consistent with being located on two
thin disks. These authors have shown that this distribution could be
reproduced by assuming that the hypervelocity stars originate from the
tidal disruption of tight binaries initially located on two stellar
disks near the Galactic center. In the tidal disruption scenario, one
component of a compact binary is ejected at a velocity comparable to
its orbital velocity around the binary's center of mass, while the
other component remains bound to the central black hole, with
eccentricity and inclination very close to those of the binary's
center of mass before disruption. To produce bound B-type stars with
eccentricities comparable to those of the S-stars, the original
separation of the incoming binary (distance between both stars) would
have to be $\sim 1$ AU \citep[e.g.,][]{gq03}. Producing unbound B-type
stars with ejection velocities $\sim 2000\;{\rm km\;s}^{-1}$ would
demand an original separation typically $\lesssim 0.1$ AU
\citep{Hills88, Bromley06}.

The presence of a couple of compact binary stars has been confirmed in
the Galactic center \citep[e.g.,][]{Martins06}. Several processes may
account for their formation. The dynamic relaxation of a population of
embedded single stars may lead to their capture into binary-star
systems. Alternatively, binary stars may form during the fragmentation
of a self-gravitating disk \citep{Alexander08}. But, these new born
binaries generally have wide separations, and they are easily
disrupted at large distance from the central black hole. The ejection
speed of such disrupted binaries would generally be too small to
produce either the hypervelocity stars or the large eccentricities of
the S-stars. To do so, additional processes that can harden the
binaries and drive their inward migration should be at work. Several
stellar dynamical processes have been considered to bring the
necessary binaries to the proximity of the SMBH \citep{perets07,
  loeckmann08, madigan09, perets10}.

Interestingly, the S-stars have many properties analogous to the
observed exoplanets, in particular the so-called Hot Jupiters. The
latter are planets typically as massive as Jupiter, orbiting within
$0.5$ AU from their host star. Some of these planets have
eccentricities up to 0.97 \citep[see e.g.,][for a recent review on
exoplanets properties]{santos08}. According to planet formation
theories, it is very unlikely to form a giant gaseous planet at such
small distances from a star \citep{lin96}. Instead, giant planets are
believed to form at larger separations in the protoplanetary gaseous
disk surrounding the central object. The tidal interaction between the
disk and the planet generally leads to decrease the planet's
semi-major axis \citep{w97}.  This is known as planetary
migration. The migration of low-mass planets (say, an Earth-mass
planet orbiting a solar mass star) has received considerable attention
as it may occur on timescales much shorter than the lifetime of the
protoplanetary disk \citep{tanaka2002}, potentially bringing all these
planets to the close proximity of their central host.

In the central parsec of our Galaxy, the star-to-supermassive black
hole mass ratio is close to the Earth-to-Sun mass ratio. It suggests
that ``planet-like migration'' could be an efficient way to bring
young stars to the inner arcsecond \citep[e.g.,][]{alp93, Linetal94,
  chang08}.  \cite{levin07} has proposed a scenario in which the
S-stars formed in a sub-pc self-gravitating disk and migrated inwards
by interacting with the disk\footnote{Not necessarily the same disk
  that originated the OB stars observed now. The S-stars could have
  formed in a previous star formation event.}. Because disk--satellite
interactions damp the satellite's eccentricities and inclinations on
short timescales, an additional process is required to explain the
large dispersion of the S-stars eccentricities and inclinations, like
for the exoplanets.  \cite{levin07} proposed a fast resonant
relaxation process that would randomize the S-stars' eccentricity and
inclination \citep{rt96}. The presence of an intermediate-mass black
hole may also influence stellar orbits, increasing their eccentricity
and inclination considerably \citep{yll07, merritt09, gualandris09}.

Our approach is to investigate the tidal interaction between a massive
binary star and its parent gaseous disk as a mechanism to harden the
binary, and to bring it to the SMBH proximity. This migration scenario
differs from the stellar scattering scenario, where stars have
parabolic encounters with the central black hole. The outline of this
paper is the following. The physical properties of the system we model
are described in \S~\ref{sec:model}. We then discuss in
\S~\ref{sec:single} the migration timescales we expect for a single
star. The numerical set-up used in our hydrodynamical simulations is
detailed in \S~\ref{sec:nummodel}. To alleviate the resolution
requirements, we show in \S~\ref{sec:rescaledmodel} that it is
possible to treat an equivalent rescaled problem. The calculation
results obtained with the rescaled problem are presented in
\S~\ref{sec:results}. We then come back to our original problem in
\S~\ref{sec:original}, and we finally discuss our results in the
context of the massive stars near the Galactic center in
\S~\ref{sec:discu}.

% ========================
\section{Physical model}
% ========================
\label{sec:model}
The system that we study comprises a supermassive black hole of mass
$M_{\rm smbh} = 3\times 10^6\;M_{\odot}$ surrounded by a gaseous disk,
where stars are assumed to have already formed by disk
fragmentation. For simplicity, only one satellite, a binary star, is
embedded in the disk. Our aim is to investigate how disk--satellite
interactions alter the orbital elements of the binary star (orbital
migration, and possible evolution of the distance between both
stars). The case of a single-star satellite with the same mass is also
considered for comparison. The properties of the binary star and of
the disk are described in \S~\ref{sec:binarymodel} and
\S~\ref{sec:diskmodel}.

% --------------------------------------
\subsection{Binary properties}
% --------------------------------------
\label{sec:binarymodel}
We consider a binary with equal-mass stars. Both stars have a fixed
mass $M_{\star} = 15\;M_{\odot}$, akin to the star S0-2 orbiting Sgr
A* \citep{martins08}. Gas accretion onto the stars is discarded. The
satellite to primary mass ratio is $q = 2M_{\star}/M_{\rm smbh} \sim
10^{-5}$, which is about three times the Earth-to-Sun mass ratio. The
motion of the stars with respect to the binary's center of mass is
prograde. Results with a retrograde binary will also be presented in
\S~\ref{sec:retrograde}. We denote by $a$ and $e$ the semi-major axis
and eccentricity of the {\it binary's center of mass}, with respect to
the central black hole. The semi-major axis and eccentricity of the
{\it binary}, with respect to its center of mass, are denoted as
$a_{\rm bin}$ and $e_{\rm bin}$. We assume that the binary has
initially $e=0$, $a=0.1\;{\rm pc}$, $e_{\rm bin} = 0$, and $a_{\rm
  bin} = 5\times 10^{-4}\;{\rm pc} \approx 100\;{\rm AU}$. The
binary's initial semi-major axis is thus $a_{\rm bin} = 0.3R_{\rm H}$,
where $R_{\rm H} = a\;(q/3)^{1/3}$ is the binary's Hill radius.

% --------------------------------------
\subsection{Properties of the gaseous disk}
% --------------------------------------
\label{sec:diskmodel}
The properties of gaseous disks around massive black holes {\it prior
  to} star formation have been described in many studies
\citep[e.g.,][]{lp87, levin07}. Such disks are expected to be
geometrically thin, with a typical aspect ratio $h$ (ratio of the
disk's scale height $H$ to orbital radius $r$) between $0.5\%$ and
$1\%$ \citep{nc05}. But, investigating star--disk interactions
requires knowledge of the disk properties {\it after} star formation,
which are largely uncertain.  The decrease of the disk density, and
the increase of its temperature depend on the number and on the mass
of stars formed \citep[e.g.,][]{sb89, naya06, ncs07}. We assume that
the increase of the disk temperature after star formation is mild, and
we set the disk aspect ratio to $h=1\%$ at $r=0.1\;{\rm pc}$ (the
initial location of our binary-star system). This yields $R_{\rm H}
\approx 1.5 H$ at the binary's initial location, which makes the
two-dimensional approximation for the disk applicable in absence of
gas accretion \citep{gda2005}. In addition, the fact $R_{\rm H} > H$
indicates that, despite its rather small mass, the binary may open a
gap around its orbit, the depth of which depends on the disk's
turbulent viscosity around the binary's orbit. The latter is modelled
by a kinematic viscosity $\nu$. The gap-opening criterion takes the
form \citep{PP3, crida06}
\begin{equation}
  \frac{3}{4}\;h(a)\left(\frac{q}{3}\right)^{-1/3} + \frac{50\nu(a)}{q\;a^2\;\Omega(a)} \lesssim 1.
\label{gapcriterion}
\end{equation}
For our disk and binary parameters, the first term in
Eq.~(\ref{gapcriterion}) is $\approx 0.5$. Assuming $\nu = \alpha h^2
a^2 \Omega$ \citep{ss73}, the second term in Eq.~(\ref{gapcriterion})
equals $50\alpha(a) h^2(a) / q$. Values of $\alpha$ in
gravito-turbulent, critically fragmenting disks should typically range
from $10^{-2}$ to $10^{-1}$ \citep[e.g.,][]{Lin88, Gammie01, Lodato04,
  Rafikov09}. Still, the value of $\alpha$ after star formation is
uncertain, as stars modify the disk's local properties (heating by
shocks, gap-opening).  For $\alpha(a) = 10^{-2}$, the second term in
Eq.~(\ref{gapcriterion}) equals 5, suggesting the binary would slowly
open a shallow gap. But, note that its actual value depends
sensitively on $h(a)$. Given the uncertainties on both $h$ and
$\alpha$, we assume that the viscosity at the binary's location is low
enough so that the binary rapidly clears a clean gap around its
orbit. For this purpose, we take $\alpha(a) = 10^{-3}$, and the second
term in Eq.~(\ref{gapcriterion}) equals 0.5.

We now describe the properties of the unperturbed disk before the
satellite opens a gap. The disk's angular velocity $\Omega$ is
slightly sub-Keplerian, the gravitational force due to the central
object being compensated for by the centrifugal force and by the
pressure force. The temperature $T$ and surface density $\Sigma$ are
set as power-law functions of $r$: $T(r) = T_{\rm
  0.1\,pc}\,(r/{0.1\,\rm pc})^{-1}$, and $\Sigma(r) = \Sigma_{\rm
  0.1\,pc}\,(r/{\rm 0.1\,pc}) ^{-1/2}$. Assuming a mean molecular
weight $\mu = 2.4$ and an adiabatic index $\gamma=5/3$, the initial
temperature at $0.1$pc is $T_{\rm 0.1\,pc} \sim 2\times 10^3$K. The
initial surface density at $0.1$ pc, $\Sigma_{\rm 0.1\,pc}$, is a free
parameter in our study. It ranges from $\sim 20\;{\rm g\;cm}^{-2}$ to
$\sim 200\;{\rm g\;cm}^{-2}$ for $Q_{\rm 0.1\,pc}$ between $1$ and
$10$, where $Q_{\rm 0.1\,pc}$ stands for the Toomre parameter at $0.1$
pc,
\begin{equation}
  Q_{\rm 0.1\,pc} = \left[ \frac{c_{\rm s} \Omega}{\pi G \Sigma} \right]_{{\rm 0.1\,pc}},
\end{equation}
with $c_{\rm s}$ the local sound speed. For simplicity, the gas self-gravity
is neglected in our study.

If stars formed along the lines of the gravitational instability's
model, the disk is expected to be radiatively efficient in the stars
forming region. After star formation, the increase of the gas
temperature due to stellar heating, and the decrease of the gas
density could significantly decrease the disk's optical thickness
\citep[e.g.,][]{bl94}. It is therefore possible that the gas becomes
radiatively inefficient after star formation. Given again the
uncertainties on the disk properties after star formation, we assume
for simplicity that the gaseous disk is radiatively efficient, modeled
with a locally isothermal equation of state (the initial temperature
profile is maintained constant with time).

% ========================
\section{Migration timescale of a single star}
% ========================
\label{sec:single}
We briefly review in this section the different regimes of migration
resulting from the tidal interaction between a gaseous disk and a
single-star satellite. A detailed review on disk--satellite
interactions can be found in \cite{masset08}. Migration timescales are
given assuming a two-dimensional disk with surface density and
temperature profiles decreasing as $r^{-\sigma}$ and $r^{-\beta}$,
respectively. The torque $\Gamma$ exerted by the disk on the satellite
alters the satellite's semi-major axis and eccentricity. Assuming the
satellite is initially on a circular orbit, the time evolution of its
semi-major axis $a$ is given by $2BqM_{\rm smbh}\;a\;\dot{a} =
\Gamma$, where $B$ is the second Oort's constant. The satellite's
migration timescale $\tau$ is defined as $\tau = |a/\dot{a}|$.

% --------------------------------------
\subsection{Type I migration}
% --------------------------------------
Type I migration applies to satellites that do not clear a gap around
their orbit. The gap-opening criterion is recalled in
Eq.~(\ref{gapcriterion}). Assuming a disk aspect ratio at the star's
location of $\sim 1\%$ (see \S~\ref{sec:diskmodel}), and a central
object of a few $10^6$ solar masses, type I migration typically
applies to stars up to a few solar masses. In this migration regime,
the torque $\Gamma_{\rm I}$ exerted by the disk on the satellite is
the sum of the differential Lindblad torque and of the horseshoe drag
\citep{pp09a}. The former accounts for the angular momentum carried
away by the spiral density waves generated by the satellite, whereas
the latter corresponds to the angular momentum exchanged between the
satellite and the gas inside its corotation region (also known as
horseshoe region). The total torque $\Gamma_{\rm I}$ can be written as
\begin{equation}
  \Gamma_{\rm I} = -C_{\rm I}q^2 \Sigma a^4 \Omega^2 h^{-2},
\label{gammaI}
\end{equation}
where all the unperturbed disk's quantities in Eq.~(\ref{gammaI}) are
to be evaluated at the satellite location. The dimensionless factor
$C_{\rm I}$ depends on $\sigma$, $\beta$, on the softening length of
the satellite's gravitational potential (required to adjust
semi-analytic estimates of the horseshoe drag with the results of
two-dimensional hydrodynamical simulations). It also depends on the
properties of the gas disk, among which its self-gravity
\citep{bm08b}, turbulent viscosity \citep{masset01, bl10}, and
radiative properties \citep{bm08a, mc09, pbck10, pbk10}. In a
two-dimensional, non self-gravitating locally isothermal disk (see
\S~\ref{sec:diskmodel}), $C_{\rm I}$ can be approximated\footnote{The
  softening length to scale height ratio is taken equal to $0.4$, and
  the horseshoe drag is assumed to be fully unsaturated. For more
  details on the horseshoe drag evaluation in isothermal disk models,
  the reader is referred e.g. to \cite{bl10}.}  as $C_{\rm I} \approx
0.8 + 1.0\sigma + 0.9\beta$ \citep[][their equation 49]{pbck10}. For
any density and temperature profiles decreasing with radius, $C_{\rm
  I}$ is positive, and type I migration is directed inwards. Using
previous notations, the timescale for type I migration, $\tau_{\rm
  I}$, reads
\begin{eqnarray}
  \left.
  \right. && \tau_{\rm I} \approx 4.3\times 10^{7} {\rm yrs}\times(0.8+1.0\sigma+0.9\beta)^{-1}\left( \frac{q}{3\times 10^{-6}} \right)^{-1} \nonumber\\
  &\times& \left( \frac{h}{10^{-2}} \right)\left( \frac{Q_{\rm 0.1\;pc}}{30} \right)\left( \frac{M_{\rm smbh}}{3\times 10^{6}\;M_{\odot}} \right)^{-\frac{1}{2}}\left( \frac{a}{\rm 0.1\;pc} \right)^{\sigma-\frac{\beta}{2}}.
\label{tauI}
\end{eqnarray}
In the particular case where $\sigma=2$ and $\beta=1$,
Eq.~(\ref{tauI}) has the same dependence with the disk and satellite
parameters as in equation (40) of \cite{levin07}. With our model
parameters ($\sigma=1/2$, $\beta=1$, $h=10^{-2}$, $M_{\rm smbh} =
3\times 10^6\;M_{\odot}$ and $a = 0.1$ pc), and further assuming $q =
10^{-7}$ ($R_{\rm H} \approx 0.3H)$ and $Q_{\rm 0.1\;pc} = 3$, we find
$\tau_{\rm I} \sim 6\times 10^7$ yrs $\approx 3.5\times 10^{4}$
orbital periods at $0.1\;{\rm pc}$. This timescale is slightly longer
than the lifetime of the stars near the Galactic center. Note that
$\sigma=1/2$ and $\beta=1$ make $\tau_{\rm I}$ independent of
$a$. With $\sigma=2$ and $\beta=1$, $\tau_{\rm I} \propto a^{3/2}$ and
goes down to a few $10^6$ yrs at $a\sim 0.01$ pc. Note also that,
since $\tau_{\rm I} \propto q^{-1}$, more massive stars (but still
subject to type I migration) will migrate faster. However, based on
our thin disk assumption, the S-stars would open a gap around their
orbit. They are therefore more likely subject to the slower type II
migration. Still, it is possible that the S-stars have substantially
migrated in before they acquired their asymptotic mass.

% --------------------------------------
\subsection{Type II migration}
% --------------------------------------
\label{sec:typeII}
Type II migration concerns satellites that are massive enough to
progressively deplete their horseshoe region, thereby opening a gap
around their orbit. Assuming again a disk aspect ratio of $\sim 1\%$,
and a central object of a few $10^6\;M_{\odot}$, type II migration
typically applies to stars more massive than $\sim
10\;M_{\odot}$. Compared to the type I migration regime, the horseshoe
drag is much reduced, and the Lindblad torque balances the viscous
torque exerted by the disk. The total torque on the satellite can be
written as a fraction $C_{\rm II}$ of the viscous torque due to the
outer disk \citep[][their equation 15]{cm07}. The factor $C_{\rm II}$
features the time-dependent fraction of gas $f_{\rm gas}$ left in the
satellite's horseshoe region.

The particular case where $f_{\rm gas}$ tends to zero is usually
called the standard type II migration regime. Its timescale reads
\begin{equation}
  \tau_{\rm II} \approx \frac{2r_{\rm o}^2}{3\nu(r_{\rm o})} 
  \left( 1 + \frac{M_{\rm sat}}{4\pi \Sigma(r_{\rm o}) r_{\rm o}^2} \right),
  \label{tauII_a}	 
\end{equation}
where $M_{\rm sat} = qM_{\rm smbh}$ denotes the satellite mass, and
where $\Sigma$ is the density of the disk perturbed by the
satellite. In Eq.~(\ref{tauII_a}), $r_{\rm o}$ is the location in the
outer disk where most of the satellite's angular momentum is
deposited. It can be approximated as the location of the outer
separatrix of the satellite's horseshoe region, given by $r_{\rm o}
\approx a + 2.5 R_{\rm H}$ for gap-opening satellites \citep{mak2006}.
  
When the satellite mass is much smaller than the local disk mass $4\pi
\Sigma(r_{\rm o}) r_{\rm o}^2$, the type II migration timescale
corresponds to the disk's local viscous timescale \citep{lp86}. In
this regime, known as disk-dominated type II migration, the migration
timescale $\tau_{\rm II, d}$ can be written as
\begin{eqnarray}
  \left. \tau_{\rm II, d}
  \right. & \approx & 1.8\times 10^{8} {\rm yrs}\times\left( \frac{\alpha}{10^{-2}} \right)^{-1}\left( \frac{h}{10^{-2}} \right)^{-2} \nonumber\\
  &\times&\left( \frac{M_{\rm smbh}}{3\times 10^{6}\;M_{\odot}} \right)^{-1/2}\left( \frac{r_{\rm o}}{\rm 0.1\;pc} \right)^{3/2},
  \label{tauII}
\end{eqnarray}
where $\alpha$ and $h$ are to be evaluated at $r_{\rm o}$. This
timescale is typically one order of magnitude longer than the
timescale for type I migration, given in Eq.~(\ref{tauI}).

In the opposite case where the satellite mass is large compared to the
local disk mass (satellite-dominated type II migration), the migration
timescale $\tau_{\rm II, s}$ is
  \begin{equation}
    \tau_{\rm II, s} \approx \tau_{\rm II, d} \times \left( \frac{M_{\rm sat}}{4\pi \Sigma(r_{\rm o}) r_{\rm o}^2} \right),
\label{tauIIb}
\end{equation}
where the satellite to local disk mass ratio is
\begin{eqnarray}
  \left. \frac{M_{\rm sat}}{4\pi \Sigma(r_{\rm o}) r_{\rm o}^2} 
  \right. & \approx & 1.7\times 10^{-4} 
  \left( \frac{M_{\rm sat}}{10\;M_{\odot}} \right) \nonumber\\
  &\times&\left( \frac{\Sigma(r_{\rm o})}{10^2\;{\rm g\;cm}^{-2}} \right)^{-1}
  \left( \frac{r_{\rm o}}{\rm 0.1\;pc} \right)^{-2}.
  \label{msat}
\end{eqnarray}
Eq.~(\ref{msat}) shows that in the early stages of their formation and
evolution, most of the massive stars near the Galactic center should
be subject to the disk-dominated type II migration, and thus migrate
in about the disk's viscous timescale.  Depletion of the gas disk, or
substantial migration towards the central object, should however slow
down migration as the stars inertial mass becomes comparable to the
local disk mass.

% --------------------------------------
\subsection{Type III migration}
% --------------------------------------
\label{sec:typeIII} Migrating satellites that open a partial gap
experience an additional corotation torque due to fluid elements
flowing across the horseshoe region (moving from the inner disk to the
outer disk if the satellite migrates inwards). At small migration
rates $\dot{a}$, this additional corotation torque is proportional to
$\dot{a}$ \citep{mp03}. When its amplitude is large enough, this
torque may thus trigger a runaway migration. Runaway occurs, roughly
speaking, when the mass deficit of the horseshoe region (the
difference between the mass of the horseshoe region, and the mass that
it would have if it had a uniform surface density equal to the density
of the orbit-crossing flow) exceeds the satellite's mass
\citep{mp03}. The occurrence and timescale of runaway migration, also
referred to as type III migration, depend sensitively on the
parameters entering the gap-opening criterion ($q$, $h$ and $\alpha$
at the satellite's location), and on the disk mass \citep{mp03}. It
concerns intermediate-mass satellites (massive satellites clear a
wide, deep gap, and the density of the orbit-crossing flow is too
small to induce a significant mass deficit) in massive disks (the more
massive the disk, the larger the density of the orbit-crossing
flow). This migration regime could then be particularly relevant to
$\gtrsim 3-10\;M_{\odot}$ stars embedded in thin ($h \sim 1\%$)
massive ($Q \lesssim 10$) disks. Numerical simulations of
disk--satellite interactions show that, depending on the resolution of
the gas flow surrounding the satellite, the timescale for runaway
migration may be as short as a few $10^2$ orbits
\citep[e.g.,][]{mp03,cbkm09}. It is thus possible that some of the
massive stars near the Galactic center formed at separations $\geq
0.1$ pc, and reached the proximity of the central black hole through
type III runaway migration.

% ========================
\section{Numerical model}
% ========================
\label{sec:nummodel}

% --------------------------------------
\subsection{Numerical method}
% --------------------------------------
All our calculations are performed with the two-dimensional code
FARGO. It is a staggered mesh code that solves the Navier-Stokes and
continuity equations on a polar grid. An upwind transport scheme is
used along with a harmonic, second-order slope limiter
\citep{vl77}. Its main feature is to use a change of rotating frame on
each ring of the grid, which increases the timestep significantly
\citep{fargo1}.

Our calculation results are expressed in the following
units. The initial semi-major axis $a_0$ of the binary's center of
mass (or of the satellite if single) is the length unit, the mass of
the central object $M_{\rm smbh}$ is the mass unit, and $(GM_{\rm
  smbh} / a_0^3)^{-1/2}$ is the time unit, $G$ being the gravitational
constant ($G = 1$ in our unit system). Whenever time is expressed in
orbital periods, it refers to as the orbital period of the binary's
center of mass (or of the satellite if single) around the central black hole.

We describe in this paragraph the numerical set-up of our
calculations. Unless otherwise stated, the disk and satellite
parameters are those described in \S~\ref{sec:model}. Since our model
assumes a locally isothermal disk, simulations do not include an
energy equation (the temperature's initial radial profile does not
evolve with time). All calculations are performed in the frame
corotating with the binary's center of mass, or with the satellite if
single. Wave-killing zones are used next to the grid's inner and outer
edges in order to minimize unphysical wave reflexions
\citep{valborro06}. In simulations with a binary satellite, we follow
\cite{Cresswell06} and set the timestep as the minimum value between
the hydrodynamical timestep, and a fixed fraction of the binary's
internal orbital period,
\begin{equation}
  \delta t_{\rm bin} = \frac{2\pi}{400}\;\left[ \frac{d_{\rm bin}^3}{2GM_{\star}} \right]^{1/2},
\end{equation}
where $d_{\rm bin}$ stands for the separation between the two
stars. To avoid a violent initial evolution of the disk, the stars'
gravitational potential is slowly turned on over 10 orbital
periods. Discussion on the form of the stars potential, and on the
grid resolution follows in \S~\ref{sec:resissue}.

% --------------------------------------
\subsection{Resolution issues}
% --------------------------------------
\label{sec:resissue}
The time-evolution of the binary should sensitively depend on how the
gas flow around and between the stars is resolved. As is usual in
disk--satellite 2D hydrodynamical simulations, the stars potential is
smoothed over a softening length $\varepsilon$. Two parameters
therefore determine the resolution of our calculations: the grid's
resolution and the softening length of the stars potential. In this
study, we choose a rather large value for $\varepsilon$ in order to
prevent a large accumulation of gas around each star, which would
otherwise require a prohibitively high grid's resolution to be
properly handled. The default value for $\varepsilon$ is half the
binary's initial semi-major axis $a_{\rm bin,0}$ (the same value is
used in simulations with a single satellite). The softening length is
therefore kept constant throughout the simulation. Smaller values of
$\varepsilon$ will be considered in \S~\ref{sec:parameterspace}.

We aim to resolve the satellite's Hill radius with about 50 cells
along the radial and azimuthal directions \citep{cbkm09}. This
requirement implies a maximum size for the grid cells equal to
$\delta_{\rm max} = 0.02 R_{\rm H} \approx 3\times 10^{-4}a$. To model
the global disk--satellite interaction, it is enough to set the grid's
radial extent to $\Delta r = 15 R_{\rm H}$. This sets the number of
cells along the radial direction to $N_{\rm r} = 750$. The cell number
along the azimuthal direction is then $N_{\rm s} = 2\pi a /
\delta_{\rm max} \approx 21000$. In contrast to $N_{\rm r}$, the value
of $N_{\rm s}$ is about one order of magnitude larger than its typical
value in 2D hydrodynamical simulations of disk--satellite
interactions. To maintain a reasonable computational cost, we show in
\S~\ref{sec:rescaledmodel} that it is possible to transform our
problem into an equivalent rescaled problem, less computationally
demanding, by making use of the dimensionless parameters in the
gap-opening criterion at Eq.~(\ref{gapcriterion}).

% ========================
\section{Rescaled problem}
% ========================
\label{sec:rescaledmodel}

% TTTTTTTTTTTTTTTTTTTTTTT
\begin{deluxetable}{cccc}
  \tablecaption{Parameters of the calculations of
    \S~\ref{sec:rescaledmodel}, where the rescaling method is
    illustrated. All runs have the same values for $d_1 = q/h^{3}$ and
    $d_2 = qa^2 \Omega/\nu$, the dimensionless quantities entering the
    gap-opening criterion.\label{table:invariance}}
  \tablewidth{0.9\hsize} \tablehead{\colhead{Parameter}&\colhead{Run
      a}&\colhead{Run b}&\colhead{Run c}} \startdata
  $h$			&$1\%$			&$3\%$				&$5\%$\\[1pt]
  $q\propto h^{3}$ &$8\times 10^{-6}$	&$2.2\times 10^{-4}$	&$10^{-3}$\\[1pt]
  $\nu/a^2\Omega \propto q$	&$8\times 10^{-8}$	&$2.2\times 10^{-6}$	&$10^{-5}$\\[1pt]
  $\alpha = \nu / h^2a^2\Omega$	&$8\times 10^{-4}$	&$2.4\times 10^{-3}$	&$4\times 10^{-3}$\\[1pt]
  $R_{\rm H}/a \propto h$			&$1.4\%$			&$4.2\%$				&$6.9\%$\\[1pt]
  $r_{\rm min}, r_{\rm max}$	&$0.88, 1.16$		&$0.64, 1.48$			&$0.4, 1.8$\\[1pt]
  $N_{\rm r}, N_{\rm s}$ &$448, 6700$ &$448, 2238$ &$448, 1344$
  \enddata
\end{deluxetable}
% TTTTTTTTTTTTTTTTTTTTTTT
%FFFFFFFFFFFFFFFFF
\begin{figure*}
  \includegraphics[width=0.5\hsize]{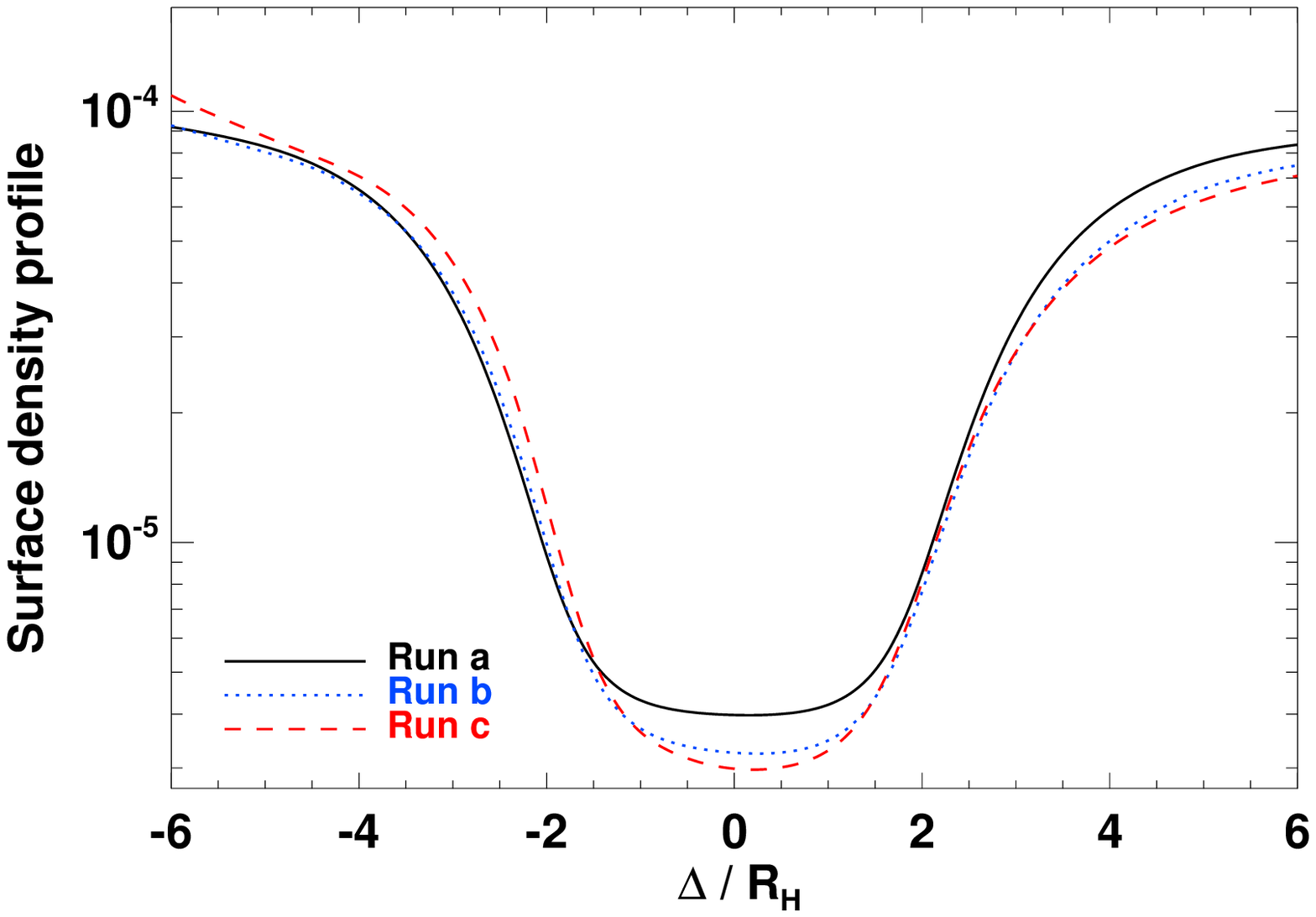}
  \includegraphics[width=0.5\hsize]{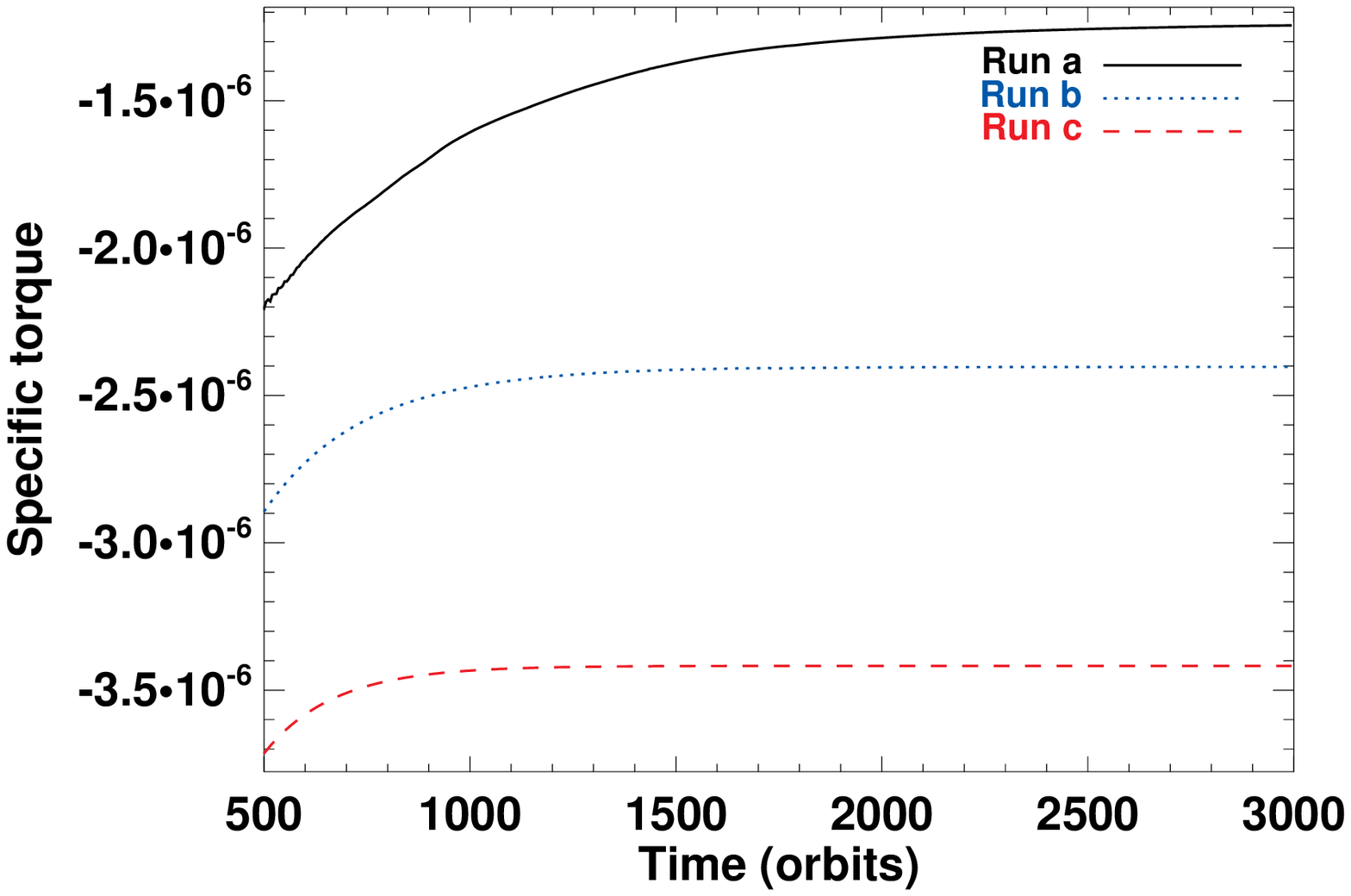}
  \caption{\label{invariance}Results of the calculations of
    \S~\ref{sec:rescaledmodel}, whose main parameters are summed up in
    Table~\ref{table:invariance}. Left: azimuthally-averaged density
    profiles at $3000$ orbits. The azimuthal averaging excludes the
    disk parts located less than a few Hill radii away from the
    satellite. The x-axis displays the disk--satellite radial distance
    $\Delta$, in units of the satellite's Hill radius. The initial
    surface density at the satellite's location equals $10^{-4}$ in
    code units. Right: time evolution of the specific torque exerted
    by the disk on the satellite.}
\end{figure*}
%FFFFFFFFFFFFFFFFF
For a single satellite with fixed semi-major axis $a$, the gap-opening
criterion involves two dimensionless quantities: $d_1 = q/h^{3}$ and
$d_2 = qa^2 \Omega/\nu$. Any changes of $q$, $h$ and $\nu$ that keep
$d_1$ and $d_2$ constant will preserve the gap's density profile
(width, depth) in scaled units $\Delta / R_{\rm H}$, where $\Delta$
stands for the disk--satellite radial distance \citep{crida06}. The
major resolution issue pointed out in \S~\ref{sec:resissue} is due to
the fact $N_{\rm s} \propto {R_{\rm H}}^{-1} \propto q^{-1/3}$. Rising
$q$, while keeping $d_1$ and $d_2$ constant, allows decreasing $N_{\rm
  s}$ while maintaining the same gap properties. We illustrate this
result for a single satellite with a series of three calculations,
referred to as Runs a, b, and c. In all three models, $d_1 = 8$ and
$d_2 = 100$. The main parameters of these runs are summarized in
Table~\ref{table:invariance}. The disk aspect ratio increases from
$h=1\%$ (Run a) to $h=5\%$ (Run c), which comes to increase the
satellite to primary mass ratio $q$ and the kinematic viscosity $\nu$
by about two orders of magnitude. All runs have $N_{\rm r} = 448$. The
location of the grid's inner and outer edges, $r_{\rm min}$ and
$r_{\rm max}$, is adjusted such that $\Delta r = r_{\rm max} - r_{\rm
  min} = 20R_{\rm H}$ ($\delta_{\rm max}/R_{\rm H} = 4.5\%$). The
number of cells along the azimuthal direction is taken as $N_{\rm s} =
4\pi a / 3 \delta_{\rm max}$, it is thus $\approx 5$ times smaller in
Run c than in Run a. In all these runs, the satellite is held on a
fixed circular orbit in order to focus on the gap properties at a
given semi-major axis. The initial disk density at the satellite
location is $10^{-4}$ in code units. Since the gas self-gravity is
neglected, and the satellite is on a fixed orbit, the gas density
value can be chosen arbitrarily.

Azimuthally-averaged density profiles at $3000$ orbits are plotted
against $\Delta/R_{\rm H}$ in the left panel of
Figure~\ref{invariance}. At a given radius, azimuthal averages exclude
the disk parts located less than a few Hill radii away from the
satellite. In scaled units $\Delta/R_{\rm H}$, the width and depth of
the gaps are in good agreement. Although not shown here, we also
checked that the density distribution inside the satellite's Hill
radius is almost indistinguishable from one calculation to
another. This is an important point to keep in mind since, as will be
shown below, the evolution of binary satellites is primarily driven by
the density distribution in the satellite's Hill radius.

We display in the right panel of Figure~\ref{invariance} the time
evolution of the specific torque (torque per unit satellite mass)
exerted by the disk on the satellite, for our series of three
runs. The specific torques differ by a factor of order unity. The
slight increase of the torque amplitude from Run a to Run c stems from
the fact (i) the surface density at the outer horseshoe separatrix,
namely at $\Delta \approx 2R_{\rm H}$, takes very similar values in
all runs, whereas (ii) the density of gas left in the gap, which
notably contributes to the positive horseshoe drag, slightly decreases
from Run a to Run c \citep[][their equation 15]{cm07}. We comment that
the timescale to reach a steady-state varies between the simulations,
which results from different gap's depletion timescales. A gap
progressively deepens as fluid elements leave the satellite's
horseshoe region. The timescale to build up a stationary gap profile
is therefore proportional to the libration period of the horseshoe
fluid elements,
\begin{equation}
  \tau_{\rm lib} \approx \frac{8\pi a}{3\Omega x_{\rm s}},
\end{equation}
where $x_{\rm s}$ is the half-width of the horseshoe region. For large
values of the dimensionless parameter $q/h^3$, $x_{\rm s} \approx
2R_{\rm H} \propto q^{1/3}$ \citep[see e.g.,][]{mak2006}. The
timescale to get a stationary gap profile thus decreases with
increasing $q$, and it should scale with $h^{-1}$. This is in good
agreement with our results. The torques differ indeed by 5\% from
their final value at about $2000$ orbits for Run a, $1000$ orbits for
Run b, and $600$ orbits for Run a. This comparison shows that, not
only our rescaling method allows reducing the simulations
computational cost\footnote{We have checked that the computational
  time scales proportional to $N_{\rm s}$, as expected.}, it also
helps to reach a faster steady-state, while getting very similar
results in terms of the density distribution and specific torque.

% ========================
\section{Results of calculations of the rescaled problem}
% ========================
\label{sec:results}
We have shown in \S~\ref{sec:rescaledmodel} that the original problem
described in \S~\ref{sec:model} can be transformed into an equivalent,
less computationally demanding rescaled problem, by using the two
dimensionless parameters in the gap-opening criterion. By equivalent,
we mean that the properties of the gap opened by the satellite (width,
depth), and thus the value of the specific torque on the satellite are
fairly preserved with the rescaling method.  This result was shown for
a single satellite. We will show in \S~\ref{sec:componebin} that the
gap properties of single and binary satellites are in good agreement,
allowing us to apply the rescaling method to binary satellites. All
the calculation results presented in this section therefore use the
rescaling method. We come back to our original problem in
\S~\ref{sec:original}.

Before presenting our results, we briefly enumerate the parameters of
the rescaled problem that we tackle in this section. The satellite to
primary mass ratio is $q = 10^{-3}$, and the disk aspect ratio at the
satellite location is $h(a) = 5\%$. The disk's kinematic viscosity is
constant, its value can be related to an equivalent alpha parameter at
the satellite location $\alpha(a) = 4\times 10^{-3}$. As shown in
\S~\ref{sec:rescaledmodel}, this set of parameters gives similar
results (gap properties, specific torque on a single satellite) to
that of our original, Galactic-center motivated problem in
\S~\ref{sec:model}: $q = 10^{-5}$, $h(a) \lesssim 1.1\%$, and
$\alpha(a) = 10^{-3}$. The initial surface density at the satellite
location is $10^{-4}$ in code units. The initial Toomre parameter at
the binary's location $\sim 160$, but (much) smaller values will be
investigated in \S~\ref{sec:parameterspace}. The grid now includes
$N_{\rm r} = 800$ rings and $N_{\rm s} = 3000$ sectors, and it extends
from $r_{\rm min}=0.55a$ to $r_{\rm max}=1.6a$ along the radial
direction. The satellite's Hill radius is resolved by about 50 cells
along the radial direction, and by about 33 cells along the azimuthal
direction. Initially, the binary's semi-major axis is $a_{\rm bin,0} =
2\times 10^{-2}\;a_0 \approx 0.3R_{\rm H}$. The softening length of
the stars potential is $\varepsilon = a_{\rm bin,0}/2$, and, to avoid
a violent initial evolution, the mass of the binary stars is slowly
increased until it reaches its proper value at 10 orbital periods.

Our calculations are performed in two steps. The satellite is first
held on a fixed circular orbit for $500$ orbits, during which the
satellite opens a gap. During this stage, the force exerted by the
disk on the satellite is ignored. Both the quantities $a$ and $a_{\rm
  bin}$ therefore remain stationary during this first stage. Because
initially $a_{\rm bin} = 0.3R_{\rm H}$, the increase of the binary's
eccentricity due to the central star remains small (on average,
$e_{\rm bin} \sim 0.04$). After $500$ orbits, the force exerted by the
disk on the satellite is included, inducing the satellite's orbital
evolution. For comparison, calculations were performed both with a
binary and a single satellites of the same $q$.

% --------------------------------------
\subsection{Binary satellite on a fixed circular orbit}
% --------------------------------------
\label{sec:componebin}
%FFFFFFFFFFFFFFFFF
\begin{figure*}
  \includegraphics[width=0.5\hsize]{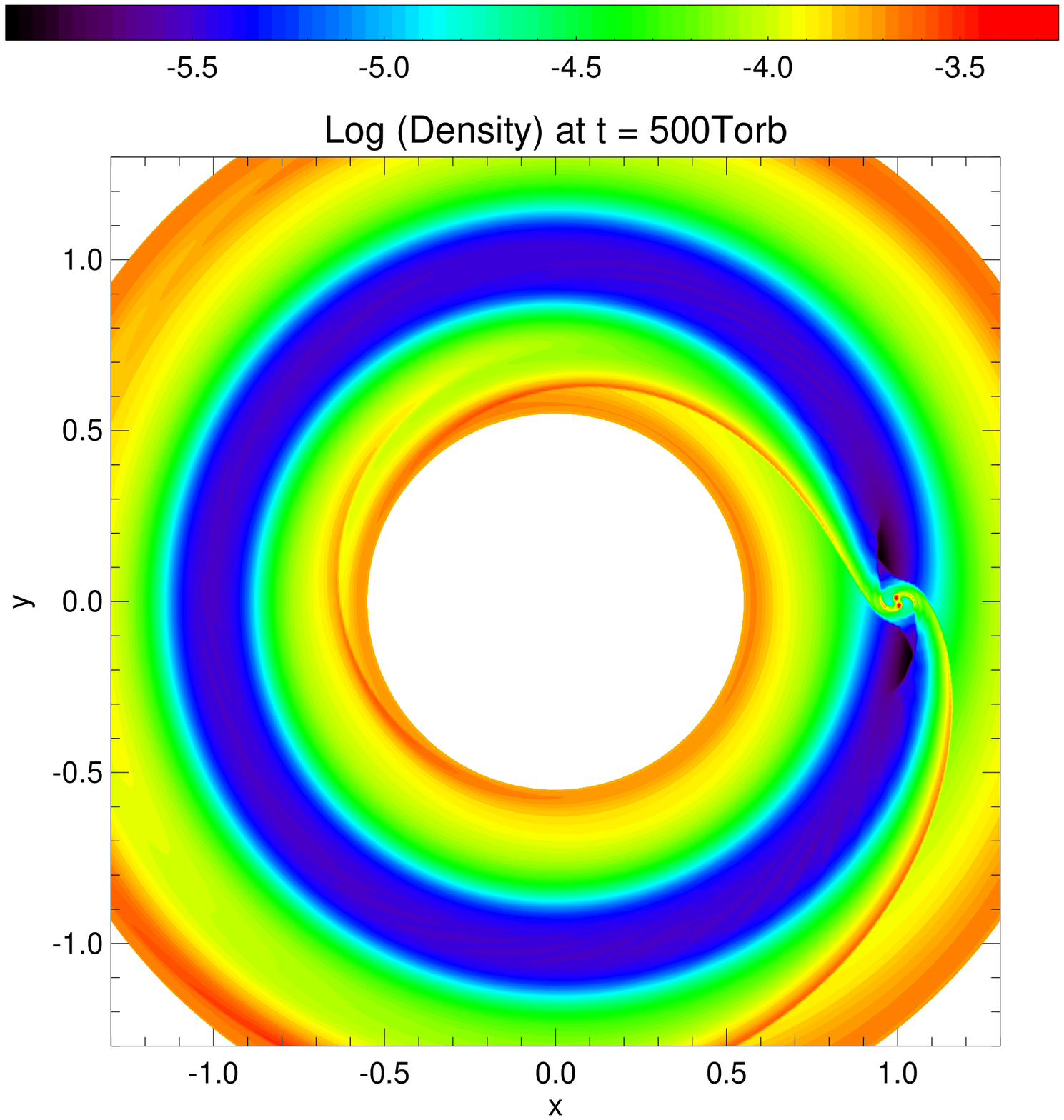}
  \includegraphics[width=0.5\hsize]{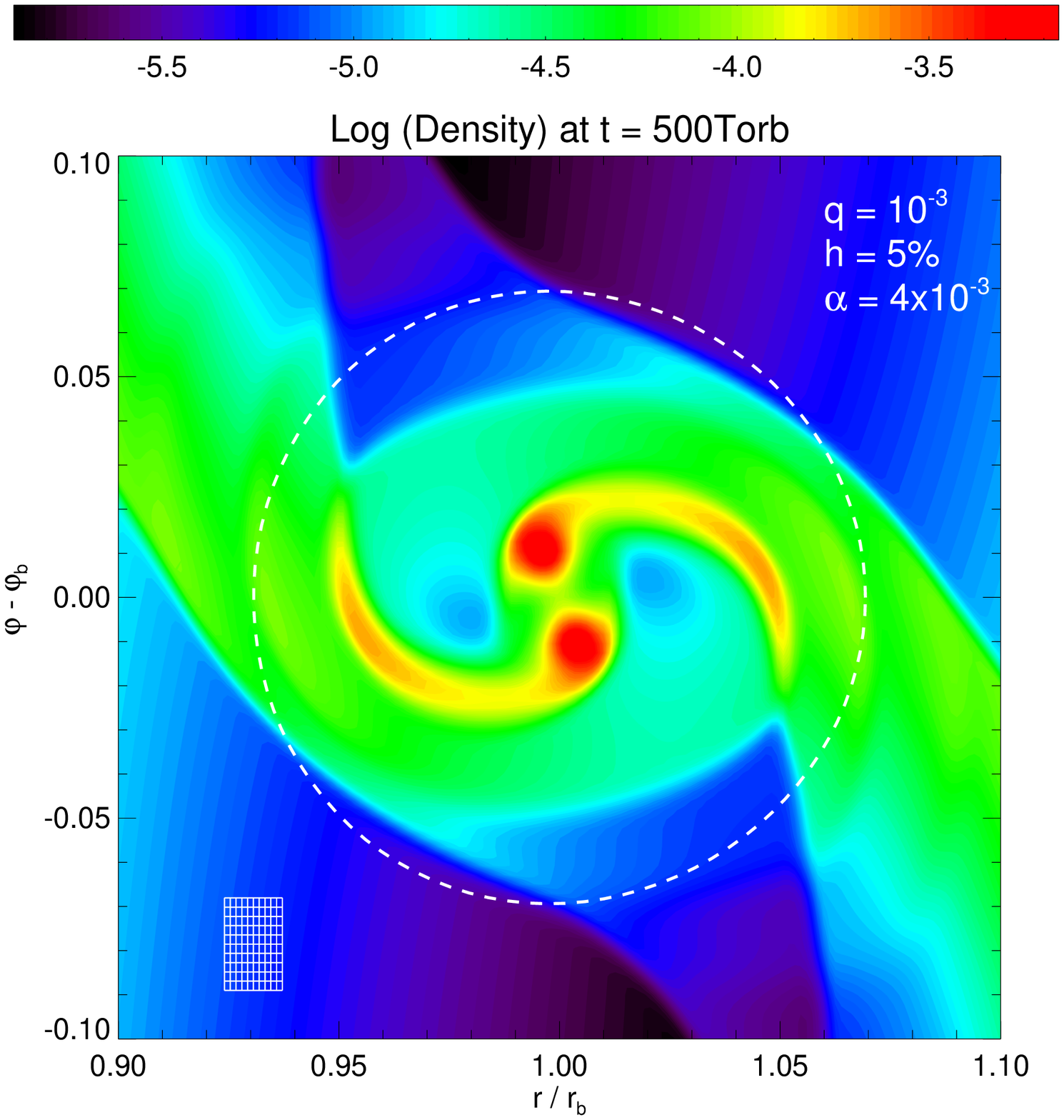}
  \includegraphics[width=\hsize]{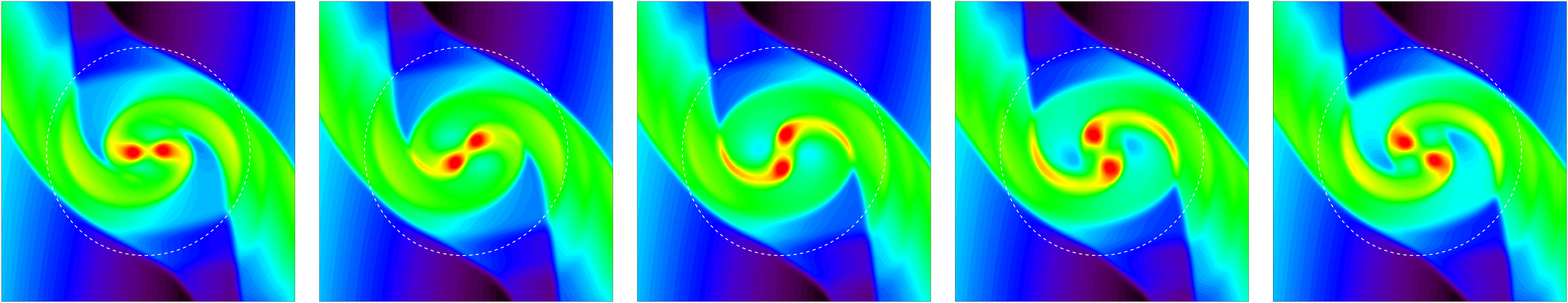}
  \caption{\label{densplot}Calculation result of the rescaled problem
    ($q=10^{-3}$, $h=5\%$, $\alpha=4\times 10^{-3}$). Top-left:
    surface density contour at $500$ orbits of the gaseous disk
    surrounding the supermassive black hole, located at $x=0$,
    $y=0$. The binary's center of mass is located at $x=1$,
    $y=0$. Top-right: close-up of the density distribution around the
    binary (note that polar coordinates are used in this panel, the
    binary's center of mass is located at $r=r_{\rm b}$, $\varphi =
    \varphi_{\rm b}$). The dashed circle shows the binary's Hill
    radius. Part of the computational grid is overplotted in the
    bottom-left part of the panel. The bottom panels show a sequential
    time evolution of the disk density in the binary's Hill radius
    over almost half a revolution of the stars around their center of
    mass (which corresponds to about 0.05 orbital periods around the
    central black hole). The color scale and the axes are the same as
    in the top-right panel.}
\end{figure*}
%FFFFFFFFFFFFFFFFF
%FFFFFFFFFFFFFFFFF
\begin{figure*}
  \includegraphics[width=0.5\hsize]{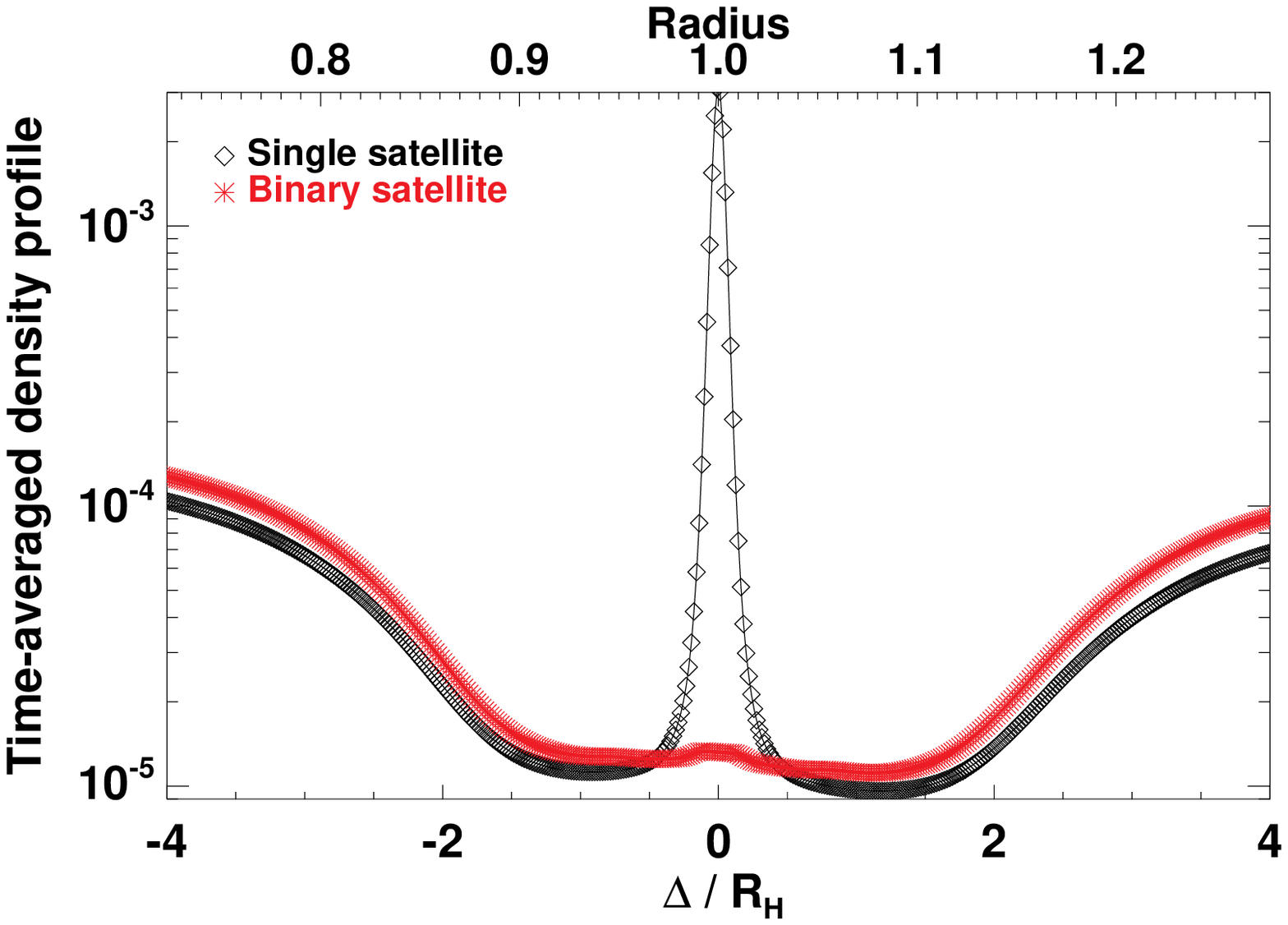}
  \includegraphics[width=0.5\hsize]{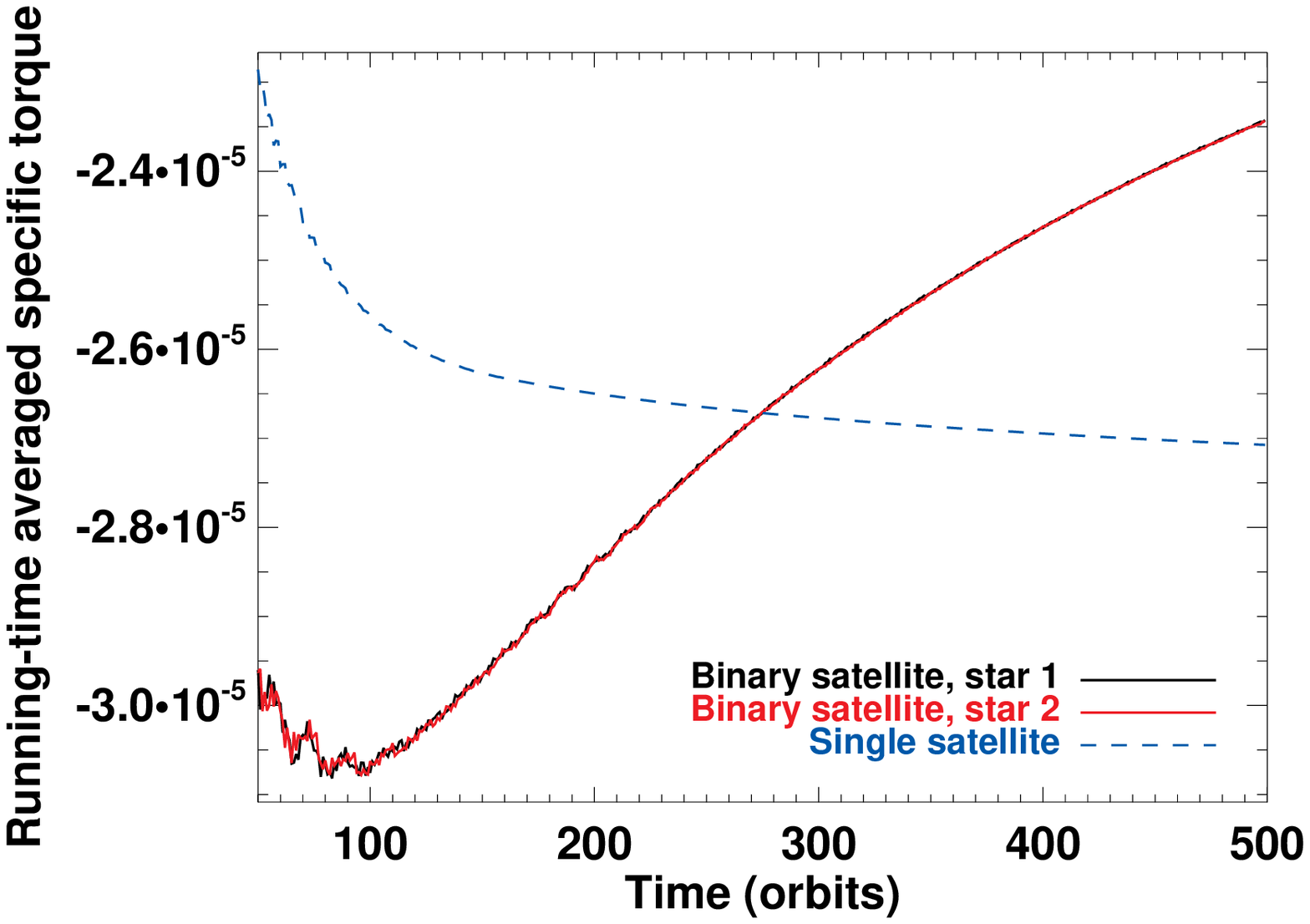}
  \caption{\label{gapopening}Left: surface density profile
    time-averaged over the $500$ orbits the satellite is held on a
    fixed circular orbit. In contrast to the left panel of
    Figure~\ref{invariance}, the azimuthal averaging takes into
    account the full azimuthal extent of the grid. The bottom x-axis
    displays the disk--satellite radial distance $\Delta$, in units of
    the satellite's Hill radius. The top x-axis shows the separation
    from the central object. Right: running-time averaged specific
    torque exerted on each satellite (the two stars in the binary
    satellite case). Time is given in orbital periods of the satellite
    around the central black hole.}
  \end{figure*}
  %FFFFFFFFFFFFFFFFF
  We describe in this paragraph the results obtained in the first
  $500$ orbits the binary satellite is held on a fixed circular
  orbit. The disk surface density obtained at $500$ orbits is
  displayed in the top-left panel of Figure~\ref{densplot}. The
  central object is located at $x=0$, $y=0$, the binary's center of
  mass is at $x=1$, $y=0$. A close-up on the density distribution
  around the binary is shown in the top-right panel of
  Figure~\ref{densplot}. The stars are located at the center of the
  two densest points in this plot. The dashed circle depicts the
  satellite's Hill radius, the size of which is $\approx
  1.4H(a)$. Part of the computational grid is displayed in the
  bottom-left part of this panel. The binary is prograde, stars thus
  rotate counterclockwise in this panel. We see that each star is
  lagged by a spiral wake (or tail). The impact of these wakes on the
  binary's orbital evolution is presented in
  \S~\ref{sec:migration}. The bottom panels in Figure~\ref{densplot}
  display a time evolution sequence of the density inside the binary's
  Hill radius during almost half a revolution of the stars around
  their center of mass (that is during $\approx 0.05$ orbital periods
  around the central black hole). The color scale and the axes are
  identical as in the top-right panel of this figure. This sequence
  shows that the density in the trailing tails is strengthened as the
  stars cross the tidal wake induced by the binary's center of mass.
  This density enhancement arises from a larger velocity difference
  between the stars and the gas at the location of the tidal wake.

  Far enough from the binary's Hill radius, the disk density much
  resembles the typical disk density obtained with a single satellite,
  as expected (the quadrupole component of the binary's potential
  becomes negligibly small compared to the monopole outside of the
  binary's Hill radius). The left panel of Figure~\ref{gapopening}
  compares the time-averaged surface density profiles obtained with
  the single and binary satellites. Time-averaging is done over the
  $500$ orbits of the simulations. In contrast to the left panel of
  Figure~\ref{invariance}, the azimuthal averaging does not discard
  the material close to the satellite's location. This is meant to
  highlight that the disk mass in the vicinity of the single satellite
  is about two orders of magnitude larger than the one around the
  binary. In the binary case, the fast relative motion of the stars
  and of the background gas hinders gas accumulation inside the
  satellite's Hill radius. This is in contrast to the single
  satellite, where a massive circum-satellite disk forms inside the
  Hill radius \citep[see e.g.,][]{cbkm09}.  Despite the large mass
  difference inside the Hill radius, the width and depth of the gaps
  opened by the single and binary satellites are in good agreement.

  The right panel of Figure~\ref{gapopening} displays the running-time
  averaged specific torque exerted on each star of the binary system
  (solid curves). Recall that time is in units of the satellite's
  orbital period around the black hole (as in all other figures). Both
  torques are indistinguishable. For comparison, the running-time
  averaged specific torque on the single satellite is also shown
  (dashed curve). At $500$ orbits, the torque obtained in the binary
  case is about $10\%$ more positive, which results from a slightly
  less depleted gap, as illustrated in the left panel of
  Figure~\ref{gapopening}. In addition, we notice from the time
  evolution of the torques that the gap's depletion timescale takes
  somewhat longer with a binary satellite. In particular, the
  running-time averaged torque on the binary has not reached a
  steady-state at $500$ orbits. We will show in
  \S~\ref{sec:parameterspace} that it does not change our results for
  the binary's orbital evolution.

% --------------------------------------  
\subsection{Binary's orbital evolution}
% --------------------------------------
\label{sec:migration} 
%FFFFFFFFFFFFFFFFF
\begin{figure}
   \includegraphics[width=\hsize]{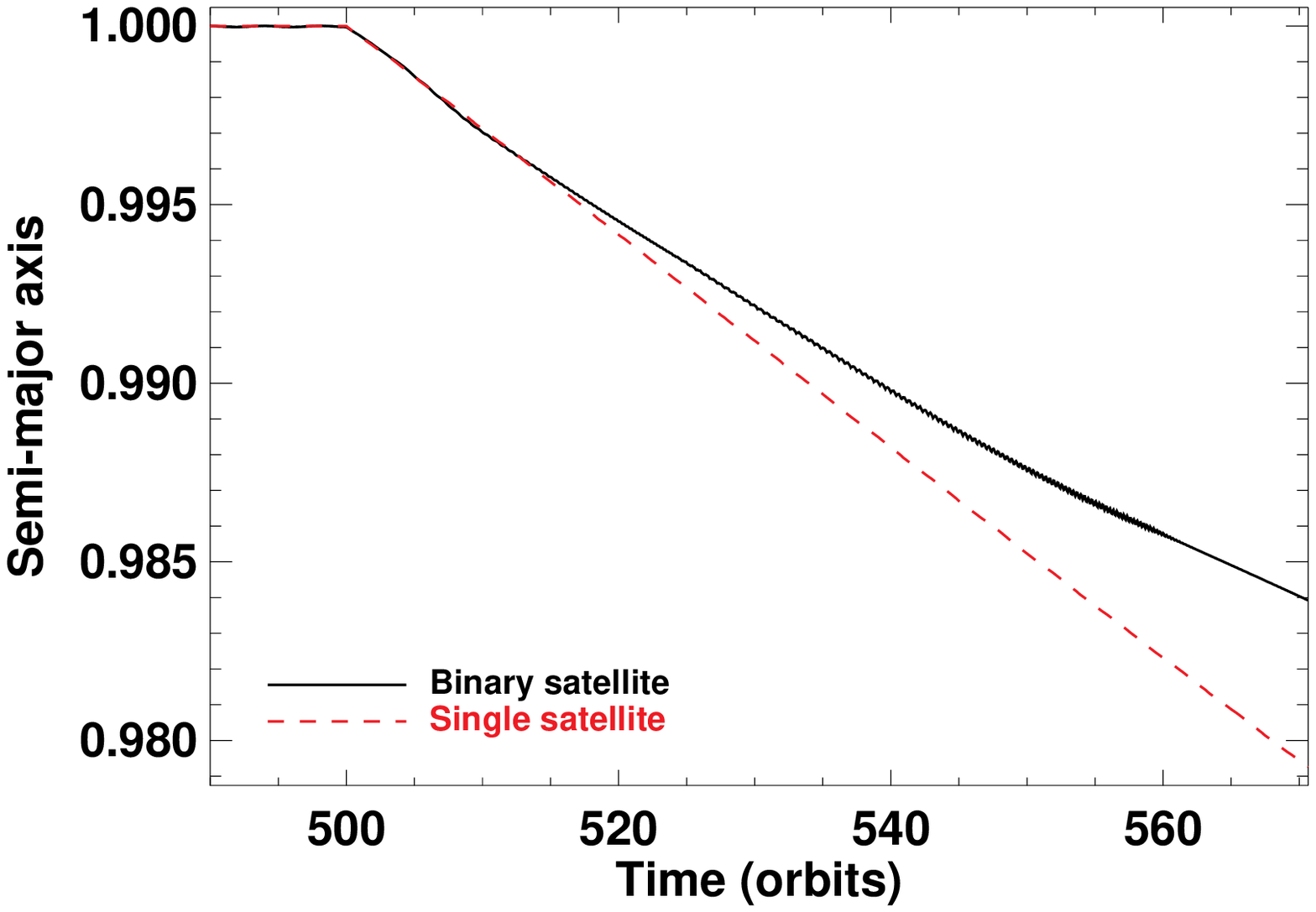}
   \includegraphics[width=\hsize]{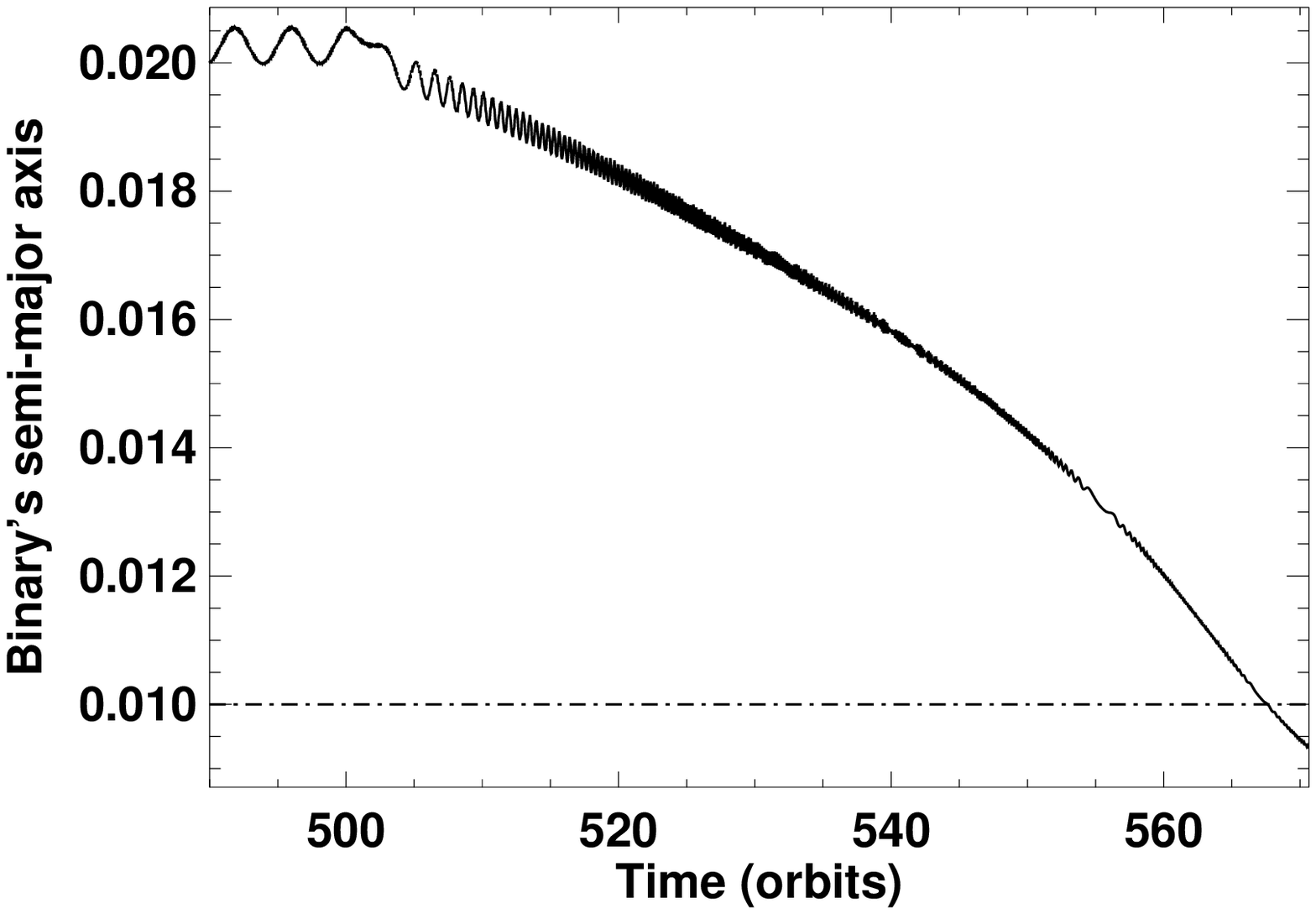}
   \includegraphics[width=\hsize]{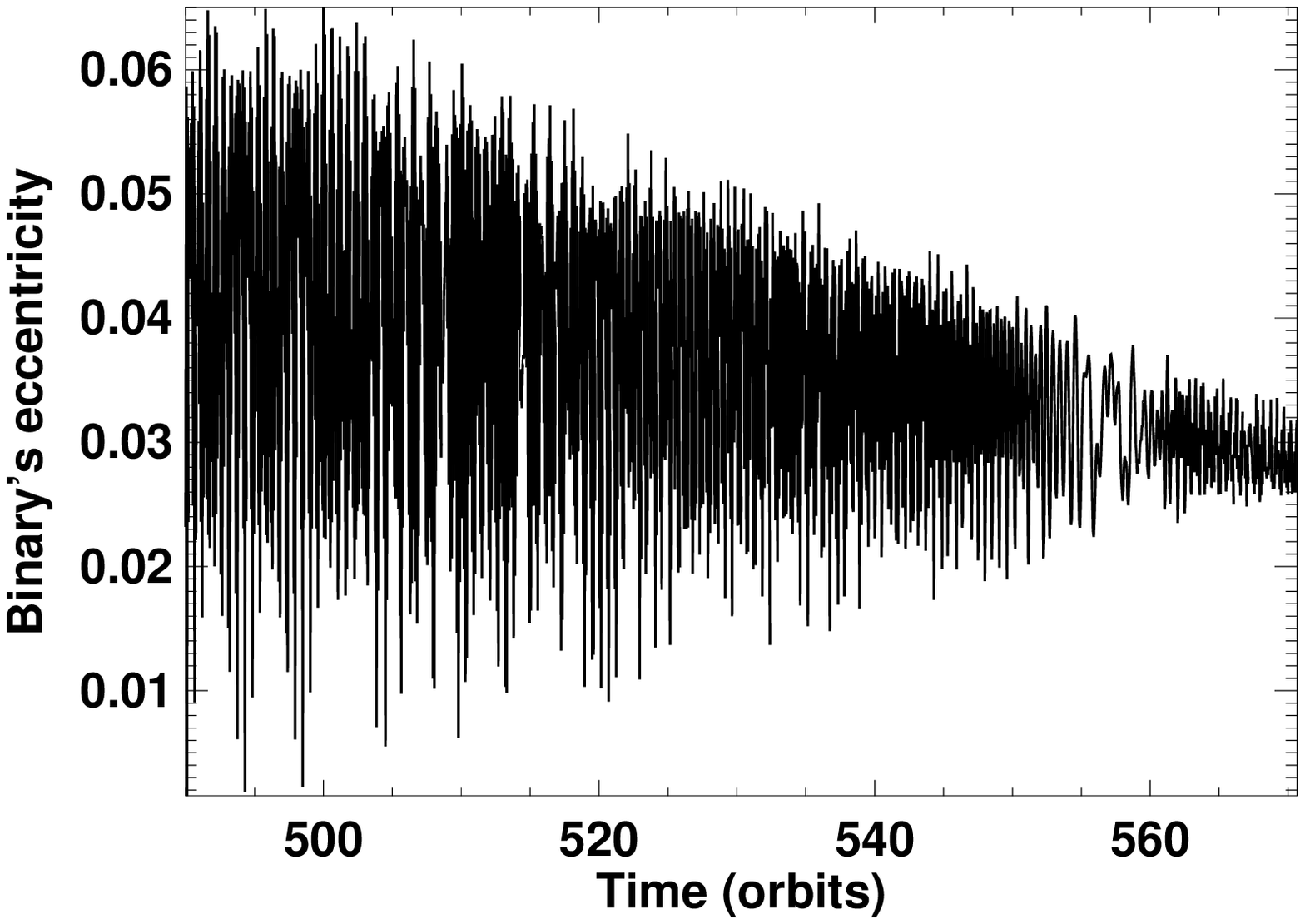}
   \caption{\label{migration}Orbital evolution of the binary star in
    its parent gaseous disk. The top panel shows the semi-major axis
    $a$ of the binary's center of mass (solid curve). The calculation
    result with a single satellite is overplotted for comparison
    (dashed curve). The binary's semi-major axis $a_{\rm bin}$ is
    displayed in the middle panel. The calculation was halted when the
    separation between the binary stars became approximately smaller
    than the stars softening length, depicted as an horizontal
    dash-dotted line in this panel. Recall that both $a$ and $a_{\rm
      bin}$ are expressed in units of $a_0$, the value of $a$ before
    restart.  The binary's eccentricity $e_{\rm bin}$ is shown in the
    bottom panel.}
\end{figure}
%FFFFFFFFFFFFFFFFF
%FFFFFFFFFFFFFFFFF
\begin{figure}
  \includegraphics[width=\hsize]{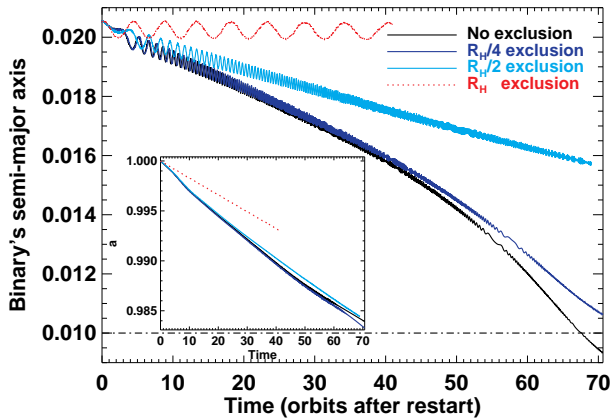}
  \caption{\label{inftronc}Impact of the disk's density distribution
    inside the binary's Hill radius on its hardening. We show the
    result of four runs restarted from the simulation of
    \S~\ref{sec:componebin}. In these four runs, the calculation of
    the force (or the torque) exerted by the disk on the binary stars
    excludes a fraction $f$ of the binary's Hill radius $R_{\rm H}$
    (centered on the binary's center of mass). The fraction $f$
    equals $0$, $1/4$, $1/2$ and $1$ from bottom to top. The
    particular case with no material exclusion ($f=0$) corresponds to
    the calculation results shown in Figure~\ref{migration}. When the
    torque exerted by all the fluid elements in the binary's Hill
    radius is discarded ($f=1$), the binary does not harden with
    time. The semi-major axis of the binary's center of mass is
    displayed in the inset plot for all four calculations.}
\end{figure}
%FFFFFFFFFFFFFFFFF
The simulation presented in \S~\ref{sec:componebin} was restarted at
$500$ orbits, allowing the binary stars to feel the force exerted by
the disk. As previously, we compare calculation results of single and
binary satellites. The semi-major axis of the binary's center of mass
(solid curve), and of the single satellite (dashed curve) are
displayed in the top panel of Figure~\ref{migration}. Both the single
and binary satellites migrate inward. The binary's migration rate is
slightly smaller, however, and it tends to decrease with time. This
behavior is consistent with the slow decrease of the running-time
averaged specific torque shown in the right panel of
Figure~\ref{gapopening}, and is due to the minor differences in the
gas density profiles depicted in the left panel of
Figure~\ref{gapopening}. Extrapolating the results in the top panel of
Figure~\ref{migration}, we find that the migration timescale of the
binary's center of mass is about $5000$ orbits, or $\sim 8\times 10^6$
yrs at $0.1$ pc. This is about a factor of $2$ shorter than the
prediction of Eq.~(\ref{tauII}), which however estimates the minimum
timescale expected in the type II migration regime. We have checked
that this discrepancy arises from the torque exerted by the
circum-satellite disk inside of the satellite's Hill radius. Another
restart run with the single satellite, wherein the calculation of the
force exerted on the satellite excludes the disk material inside of
the satellite's Hill radius, gives a migration rate that is about half
that of the simulation shown in the top panel of
Figure~\ref{migration}, and which is thus consistent with the pace
expected in the disk-dominated type II migration regime. This behavior
is further commented and illustrated for the binary satellite in the
inset panel of Figure~\ref{inftronc}.

As shown in the middle panel of Figure~\ref{migration}, the binary's
semi-major axis $a_{\rm bin}$ decreases with time. We observe that the
amplitude of the {\it hardening rate}, namely the quantity
$|\dot{a}_{\rm bin}|$, slowly increases with time. The calculation was
stopped when the separation between both stars went slightly below the
softening length $\varepsilon = a_{\rm bin,0}/2$, depicted as a
dash-dotted line in this panel. Below this separation, the binary
satellite mostly interacts with the surrounding disk like a
single-star satellite. In particular, we have checked that a massive,
circular circumbinary disk starts to form around both stars, like with
a single-star satellite. Our calculation result indicates that the
binary's semi-major axis is reduced by a factor of $2$ in $\approx 70$
orbits around the central black hole, or in about $1000$ orbits of the
stars around their center of mass. This corresponds to $\sim 10^5$ yrs
at $0.1$ pc in the Galactic center case. In this particular
simulation, the hardening timescale of the binary is $\sim 40$ times
shorter than its migration timescale.

The bottom panel of Figure~\ref{migration} shows the time evolution of
the binary's eccentricity $e_{\rm bin}$. The interaction with the gas
slowly decreases both the averaged value, and the amplitude of the
eccentricity's oscillations. Although not shown here, we have checked
that the eccentricity of the binary's center of mass remains
negligibly small with time.

The binary's hardening results from the formation of a spiral tail at
the trailing side of each star \citep{Escala04, Kim08, s10}.
\cite{Kim08} showed that each perturber is dragged backward by its own
wake, and pulled forward by the wake of its companion. Differently
said, each companion of a binary system loses angular momentum due to
its own wake, and gains angular momentum from its companion's
wake. For equal-mass perturbers, the ratio of these positive and
negative torques is always smaller than unity, and it varies with the
perturbers' velocity with respect to the background gas
\citep{Kim08}. The net effect of the wakes is therefore to extract
angular momentum from each star, which is ultimately transmitted to
the disk though viscous shear or shocks. This is analogous to the net
effect of the inner and outer wakes generated by the binary's center
of mass in the disk around the central black hole.

To further illustrate the connection between the hardening of the
binary and the flow inside its Hill radius, we performed additional
simulations wherein the calculation of the force exerted by the disk
on the stars excludes a fraction $f$ of the binary's Hill
radius. Differently said, the binary stars feel the force exerted by
all the fluid elements, except those located inside a circle centered
on the binary's center of mass and of radius $f R_{\rm H}$. We took
three values for $f$: $1/4$, $1/2$ and $1$. The results of these
simulations are displayed in Figure~\ref{inftronc}. The result with
$f=0$, which corresponds to the simulation presented in
Figure~\ref{migration}, is overplotted for comparison. The time in
$x-$axis is that after the restart (namely, after the first $500$
fixed circular orbits during which the satellite opens a gap). The
binary's hardening rate decreases with increasing $f$. When the torque
evaluation discards all of the Hill radius, the binary's hardening
rate is vanishingly small. This result clearly shows that the
hardening of the binary is caused by the gas inside its Hill
radius. In addition, the very similar results obtained with $f=0$ and
$f=1/4$ show that the hardening does not stem from the gas directly
surrounding the stars (the two densest points in the top-right panel
of Figure~\ref{densplot}), which is poorly resolved. The progressive
decrease of the hardening rate from $f=1/4$ to $f=1$ confirms that the
hardening does arise from the tails lagging the stars.

The inset plot in Figure~\ref{inftronc} depicts the time variation of
the semi-major axis of the binary's center of mass obtained with the
previous simulations. The migration rates obtained with $f=0$ and
$f=1/2$ are in very close agreement, whereas the binary's hardening
rates differ by a factor of $\sim 2$. It suggests that the hardening
of the binary has a very limited impact on its migration rate. For
$f=1$ (no hardening), the migration rate takes a smaller value, in
agreement with the findings with a single-star satellite
\citep{cbkm09}. Our results thus show that the hardening rate of the
binary is primarily triggered by the disk inside the binary's Hill
radius, whereas its migration rate is controlled by the disk outside
of its Hill radius. Our calculations do not suggest there is a
significant feedback from one rate on the other. This result does not
mean that modelling the global disk structure around the central black
hole is unnecessary. Akin to the single satellite case, the flow of
gas entering and leaving the Hill radius depends indeed on the
interaction of the satellite with the whole disk (e.g., by the capture
of fluid elements inside the satellite's horseshoe region). We finally
comment that as the binary hardens, more gas can flow inside the
binary's Hill radius, which can explain the slow increase of the
binary's hardening rates seen in the middle panel of
Figure~\ref{migration}.

% --------------------------------------
\subsection{Parameter-space study}
% --------------------------------------
\label{sec:parameterspace}
%FFFFFFFFFFFFFFFFF
\begin{figure*}
   \includegraphics[width=0.5\hsize]{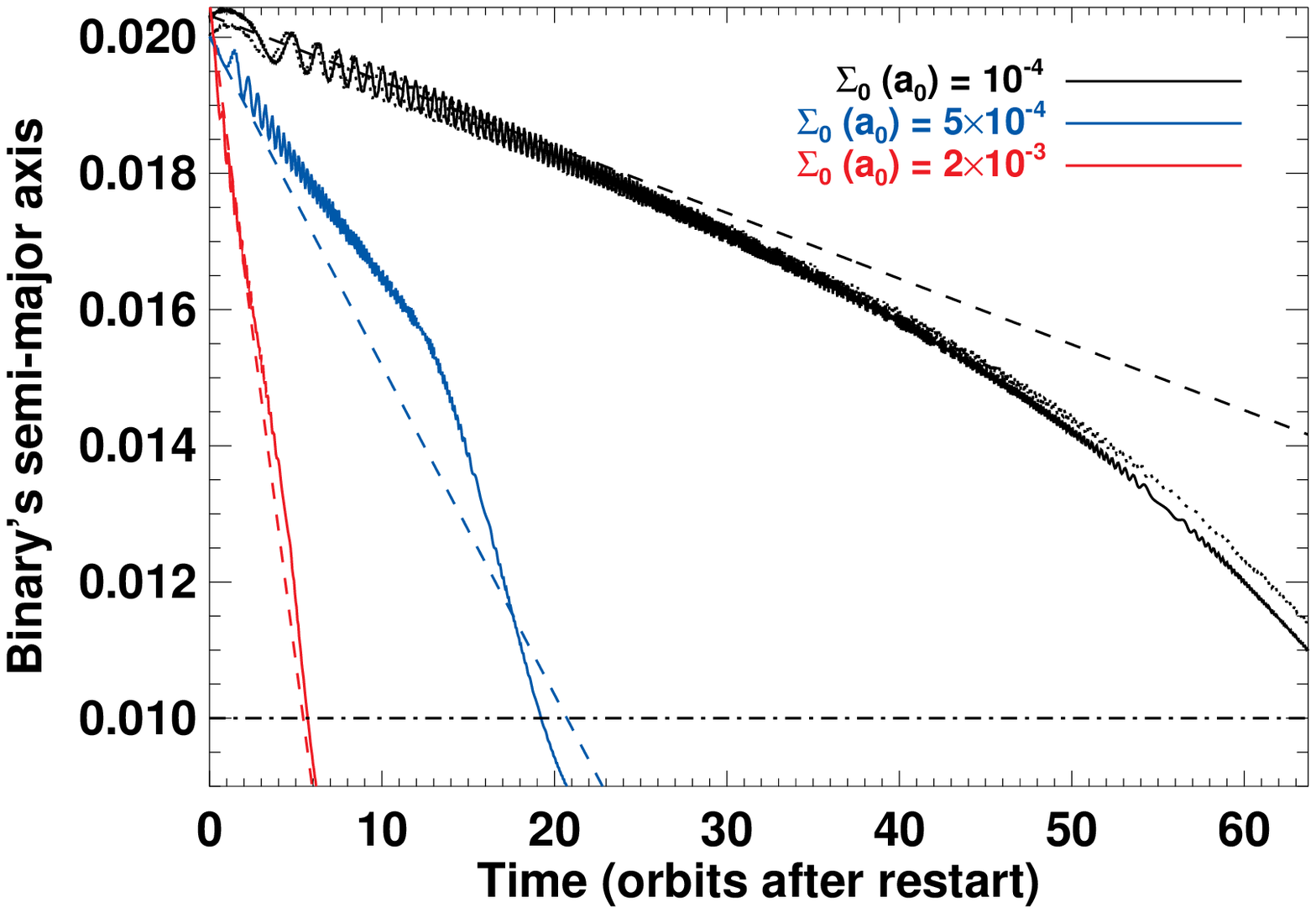}
   \includegraphics[width=0.5\hsize]{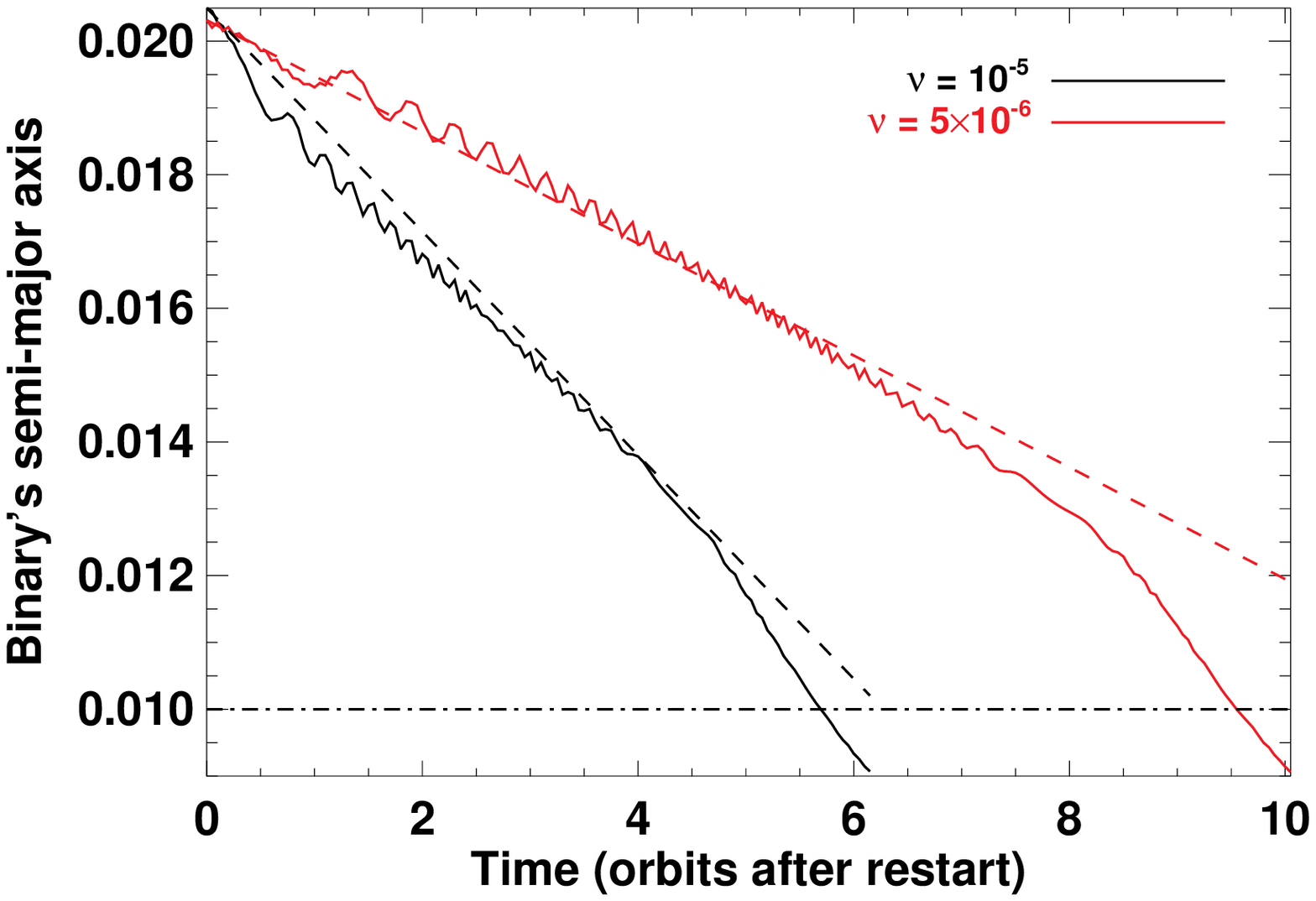}
   \includegraphics[width=0.5\hsize]{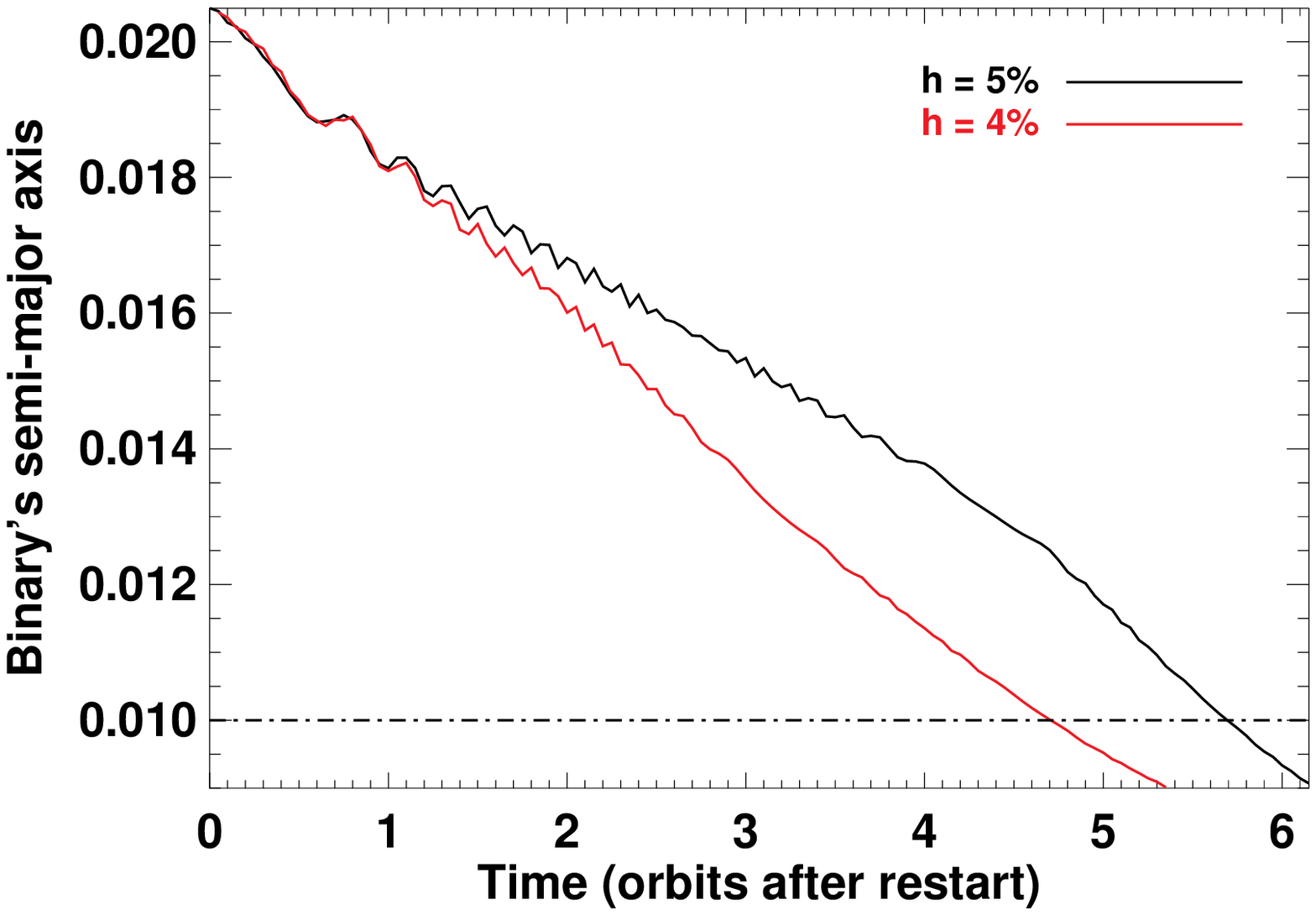}
   \includegraphics[width=0.5\hsize]{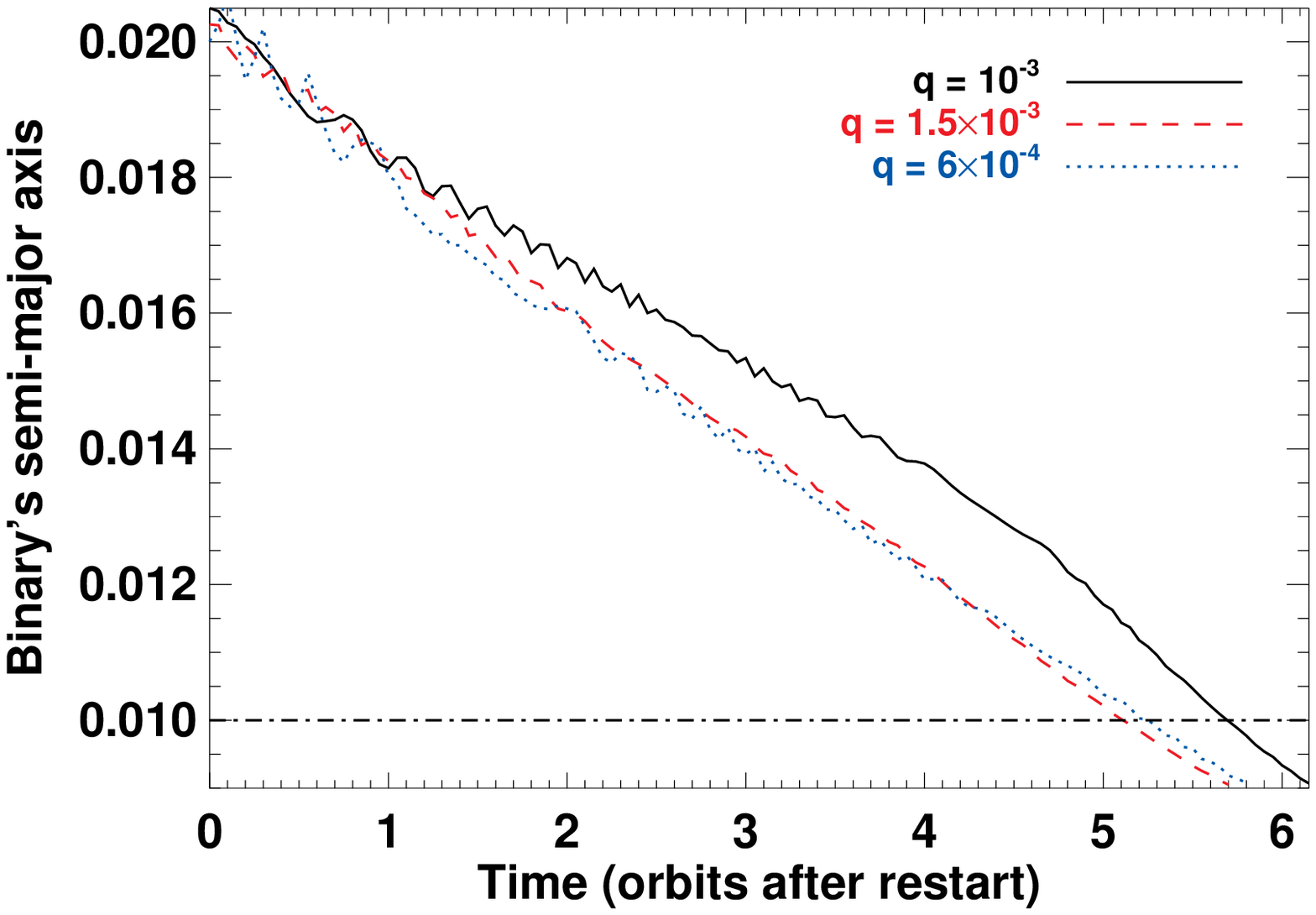}
   \caption{\label{pps}Influence of the disk and satellite parameters
     on the time evolution of the binary's semi-major axis. In all
     panels, the horizontal dash-dotted line shows the value of the
     stars softening length. {\it Top-left}: results of simulations
     with increasing the unperturbed surface density $\Sigma_0(a_0)$
     at the binary's initial location (solid curves). The black solid
     curve shows the result of the simulation in
     \S~\ref{sec:migration}. The black dotted curve shows the result
     of the same simulation restarted at $280$ orbits (instead of
     $500$ orbits). Dashed lines show linear fits to the numerical
     results, assuming the hardening rate $|\dot{a}_{\rm bin}|$ at the
     restart time scales with $\Sigma_0(a_0)$ (see text). Top-right:
     results of an additional simulation with a smaller
     viscosity. Again, dashed lines show linear fits to the numerical
     results, assuming $|\dot{a}_{\rm bin}|$ scales with $\nu$. {\it
       Bottom-left}: result of a simulation with a slightly smaller
     disk aspect ratio. {\it Bottom-right}: calculation results with
     different satellite to primary mass ratios.}
\end{figure*}
%FFFFFFFFFFFFFFFFF
%FFFFFFFFFFFFFFFFF
\begin{figure}
  \includegraphics[width=\hsize]{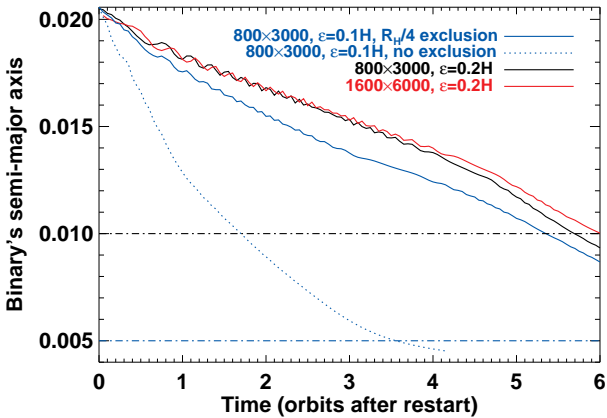}
  \caption{\label{infnum}Dependence of the binary's hardening rate
    with numerics. We show the impact of the grid's resolution at a
    fixed softening length (top solid curves), and of the softening
    length at a fixed grid's resolution. In the latter case, results
    are shown with discarding the torque of the fluid elements located
    inside $1/4$ of the binary's Hill radius (bottom solid curve), and
    without such exclusion (dotted curve). The corresponding values of
    $\varepsilon$ are depicted as horizontal dash-dotted lines.}
\end{figure}
%FFFFFFFFFFFFFFFFF
We have shown in \S~\ref{sec:migration} that the interaction between a
binary star and the gaseous disk it is embedded in leads to the
hardening of the binary as it migrates inward. The hardening timescale
is found to be much shorter than the migration timescale. This section
is aimed at investigating the dependency of the hardening timescale
upon the disk and satellite properties, and upon numerics. A detailed
exploration of the parameters space is outside the scope of this
paper. Disk and satellite parameters are allowed to vary within a
narrow range of values such that the main assumption of our model (the
binary opens a gap) is fulfilled.

We first consider the impact of the unperturbed surface density at the
binary's location. Two additional simulations were performed with
$\Sigma_0(a_0) = 5\times 10^{-4}$ and $\Sigma_0(a_0) = 2\times
10^{-3}$. The initial Toomre parameter $Q_0$ at the binary equals $32$
and $8$, respectively. The result of these simulations, and that of
the run of \S~\ref{sec:migration}, are displayed in the top-left panel
of Figure~\ref{pps} (solid curves). Increasing the unperturbed density
results in a faster hardening of the binary. For $Q_0(a_0)=8$, which
can be seen as a representative value for our original model in
\S~\ref{sec:model}, the binary's semi-major axis is reduced by a
factor of $2$ in only $\sim 5$ orbits. In that case, the mass enclosed
in the binary's Hill radius when it starts to harden is about 50 times
smaller than the binary's mass itself. The dependence of the hardening
rate with varying $\Sigma_0(a_0)$ can be interpreted as follows. As
the binary is held on a fixed orbit, and because self-gravity is
neglected, the fractional change $(\Sigma - \Sigma_0) / \Sigma_0$ of
the gas surface density is independent of $\Sigma_0$. This is true in
particular inside the binary's Hill radius, which suggests that the
net torque exerted by the trailing tails on each star, and therefore
the binary's hardening rate $|\dot{a}_{\rm bin}|$, should scale with
$\Sigma_0(a_0)$. The upper dashed curve displays a linear fit to
$a_{\rm bin}(t)$ for $\Sigma_0(a_0) = 10^{-4}$. The two bottom dashed
curves depict previous fit with a slope increased proportionally to
the unperturbed density. The good agreement with the calculation
results confirms that $|\dot{a}_{\rm bin}|$ is approximately
proportional to $\Sigma_0(a_0)$.

We have also checked the dependence of the hardening rate on the
restart time of our simulations. Restarting the run of
\S~\ref{sec:componebin} at 280 orbits gives the result overplotted in
the top-left panel of Figure~\ref{pps} as a dotted curve. It is in
very close agreement with the calculation result of
\S~\ref{sec:migration} where the restart time is $500$ orbits, despite
the fact the running-time averaged torque has not yet reached a
steady-state in either case (see right panel of
Figure~\ref{gapopening}). This agreement is consistent with the fact
that the binary's hardening is driven by the density distribution
within its Hill radius, which attains a faster steady-state than the
density distribution in the gap. We have obtained the same good
agreement with $\Sigma_0(a_0) = 2\times 10^{-3}$.

We now assess the impact of the parameters altering the gap
properties, taking $\Sigma_0(a_0) = 2\times 10^{-3}$ for
convenience. The result of calculations with varying the disk's
kinematic viscosity is displayed in the top-right panel of
Figure~\ref{pps} (the run with $\nu = 10^{-5}$, which corresponds to
$\alpha = 4\times 10^{-3}$ at the planet's location, is the same as in
the left panel of this figure). Increasing the viscosity results in a
faster hardening. Again, this result can be explained by the fact a
larger viscosity induces a larger mass flow in the satellite's Hill
radius, thereby increasing the amplitude of the drag force on each
star. As in the top-left panel of Figure~\ref{pps}, the dashed curves
highlight that the binary's hardening rate is proportional to the disk
viscosity. We have checked that this result is quantitatively
consistent with the density in the binary's Hill radius scaling
proportional to the viscosity. We have also checked that, as expected,
the migration rates obtained in these two runs scale with the disk
viscosity.

The bottom-left panel of Figure~\ref{pps} compares the results
obtained with a disk aspect ratio $h=5\%$ and $h=4\%$. Our simulations
show that the hardening is slightly faster with a smaller disk aspect
ratio, the hardening timescales differing by about $20\%$. In the same
vein, the bottom-right panel of Figure~\ref{pps} shows results with
different (albeit close) values of the satellite to primary mass
ratio, ranging from $q = 6\times 10^{-4}$ to $q = 1.5\times
10^{-3}$. These simulations have the same $a_{\rm bin,0}$. Our
calculation results show that, for the range of values that we
considered, $q$ also has little impact on the binary's hardening
rate. We have checked that the density distributions in the binary's
Hill radius show very little dependence with the aforementioned values
of $h$ and $q$, which is consistent with our findings. The gap's
depth, however, largely differs between these runs, and so do the
migration rates. We point out that in the simulation with $q = 6\times
10^{-4}$ and $h=5\%$, the binary experiences inward runaway migration,
in agreement with \cite{mp03}.
%FFFFFFFFFFFFFFFFF
\begin{figure*}
  \includegraphics[width=0.5\hsize]{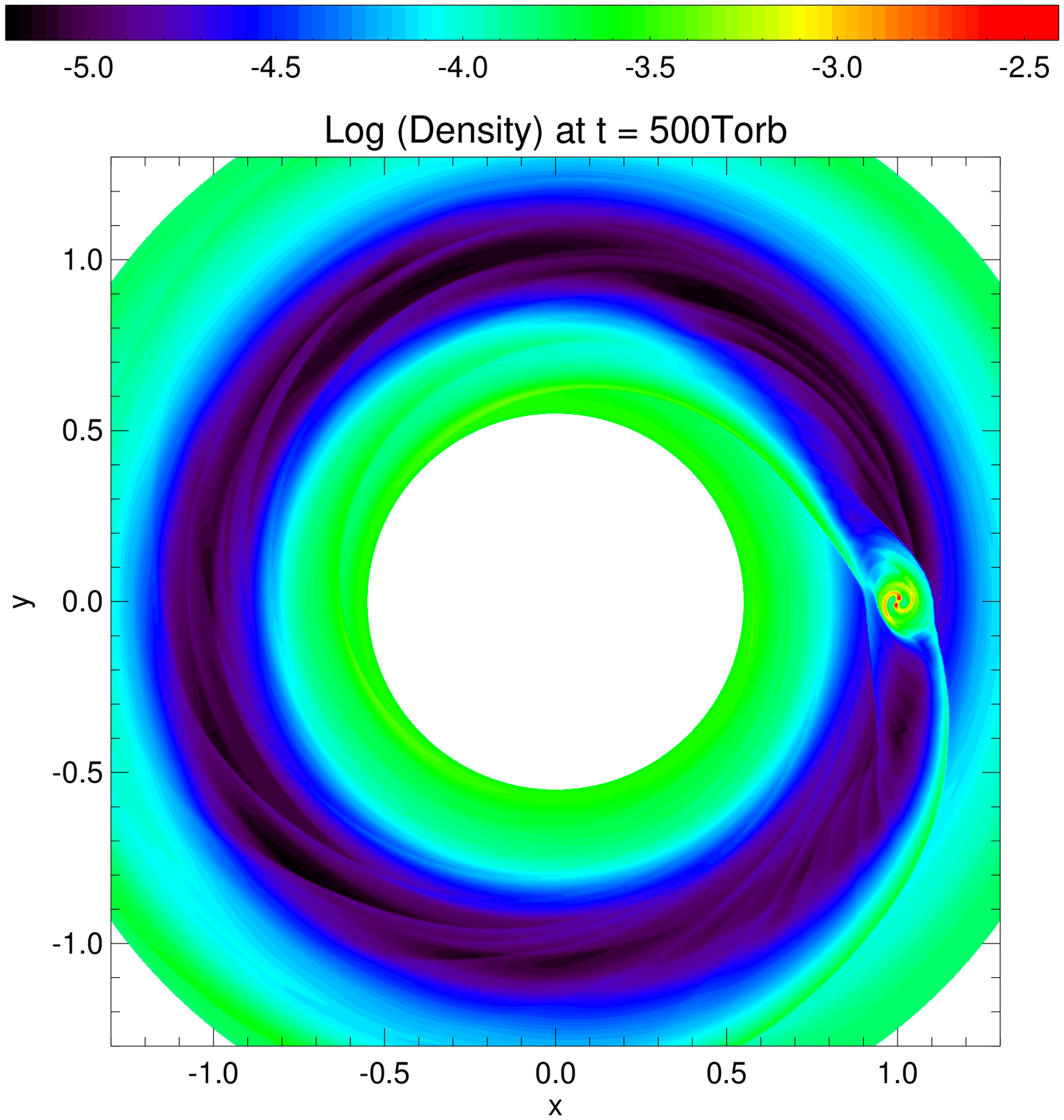}
  \includegraphics[width=0.5\hsize]{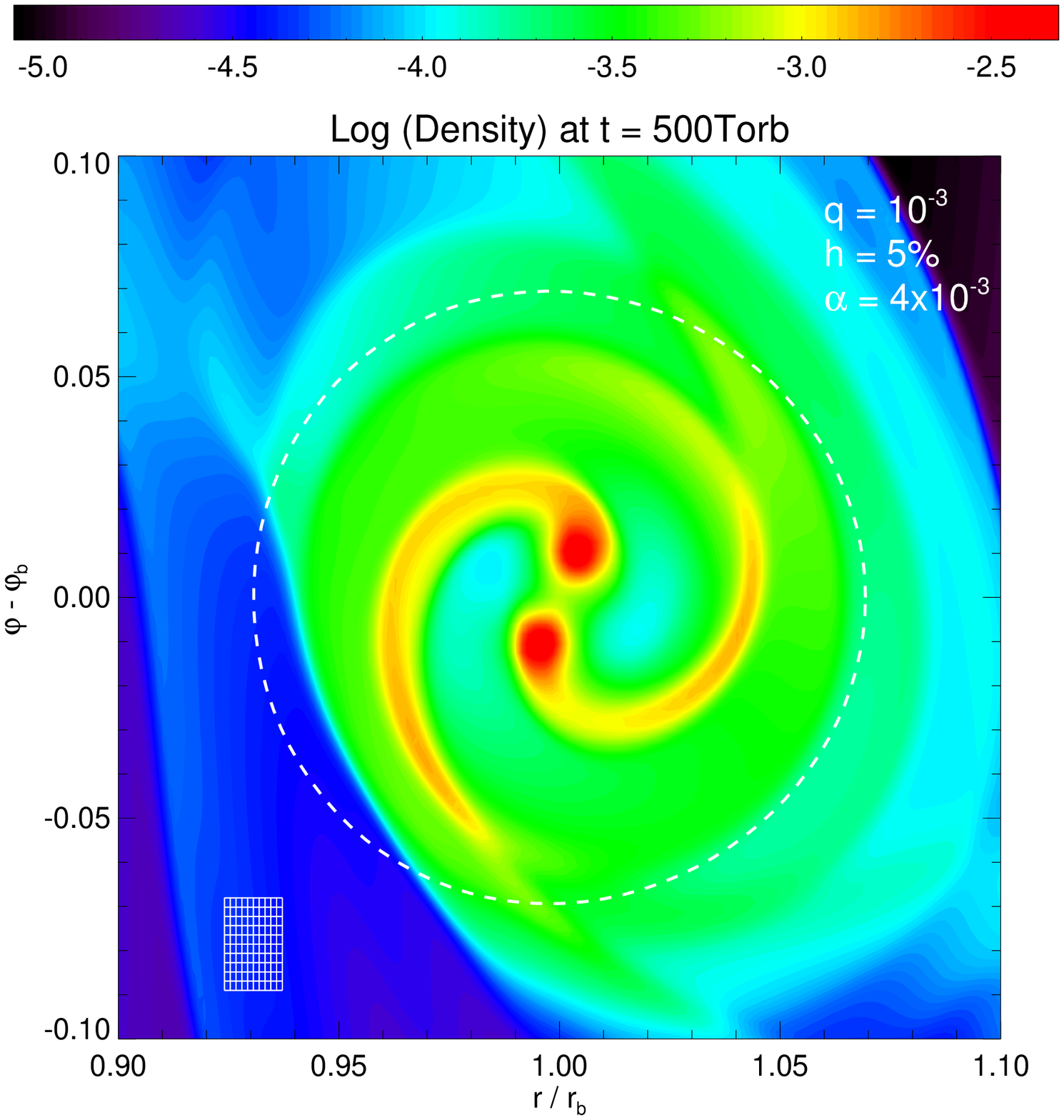}
  \caption{\label{densplotR}Left: surface density contour at $500$ orbits of
    the disk perturbed by a retrograde binary held on a fixed circular
    orbit. Right: close-up of the density distribution around the
    retrograde binary. As in Figure~\ref{densplot}, the dashed circle
    shows the binary's Hill radius, and part of the computational grid
    is overplotted in the bottom-left part of the panel.}
\end{figure*}
%FFFFFFFFFFFFFFFFF

We now discuss the dependence of our findings on the resolution of our
calculations. As already stressed in \S~\ref{sec:resissue}, resolution
is controlled by the softening length $\varepsilon$ of the stars
potential, and by the grid's resolution. We study in this paragraph
the impact of both parameters. Two additional calculations were
performed with $\Sigma_0(a_0) = 2\times 10^{-3}$: one with doubling
the number of grid cells along each direction, and one at our standard
resolution but with $\varepsilon = 0.1H$ (that is, half our standard
softening length). Results are depicted in Figure~\ref{infnum}. For
our fiducial value of $\varepsilon$, increasing the grid's resolution
is found to have no impact on the hardening rate. Decreasing the
softening length at same resolution yields however a dramatic raise of
the hardening rate (dotted curve). While the density inside the
binary's Hill radius, in particular at the location of the tails,
takes similar values for both values of $\varepsilon$, the gas density
at the immediate vicinity of the stars is increased by about one order
of magnitude for $\varepsilon = 0.1H$. It clearly suggests that the
large mass accumulation around the stars is severely under-resolved,
in contrast to the results presented with our fiducial softening
length, and that the resulting increase of the hardening rate is a
numerical artifact triggered by a lack of resolution. To illustrate
this statement, we performed an additional simulation at $\varepsilon
= 0.1H$, wherein the binary's time-evolution discards the torque
exerted by the fluid elements located inside a circle of radius
$R_{\rm H}/4$ centered on the binary's center of mass. As already
stated in \S~\ref{sec:migration}, such radius is meant to discard the
torque due to the mass accumulation around the stars, with very small
impact on the torque exerted by the tails (see
Figure~\ref{inftronc}). The result of this simulation is overplotted
in the bottom-right panel of Figure~\ref{infnum}, and shows a good
agreement with the result at larger softening.

% --------------------------------------  
\subsection{Case of a retrograde binary}
% --------------------------------------
\label{sec:retrograde}
%FFFFFFFFFFFFFFFFF
\begin{figure*}
  \includegraphics[width=0.5\hsize]{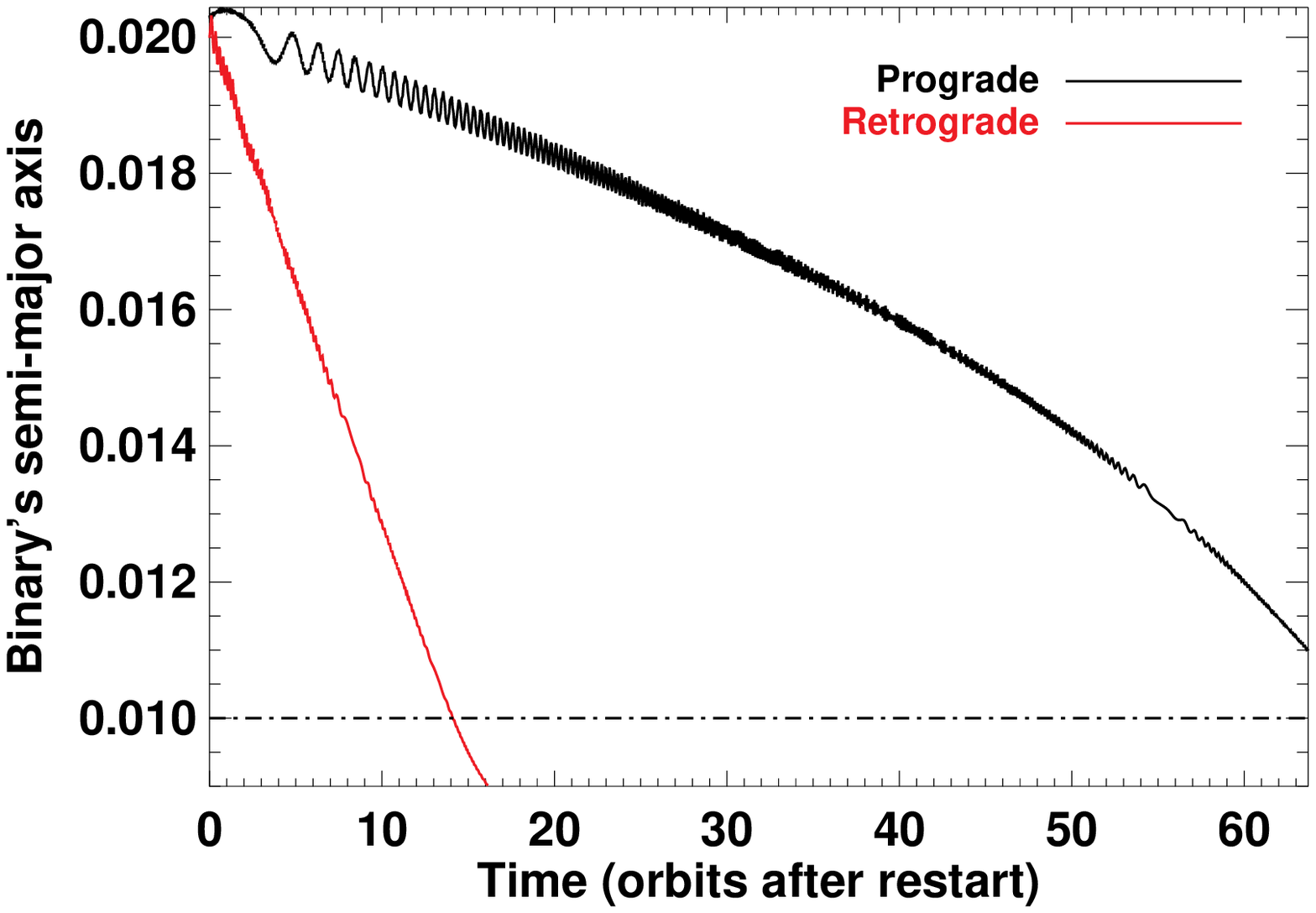}
  \includegraphics[width=0.5\hsize]{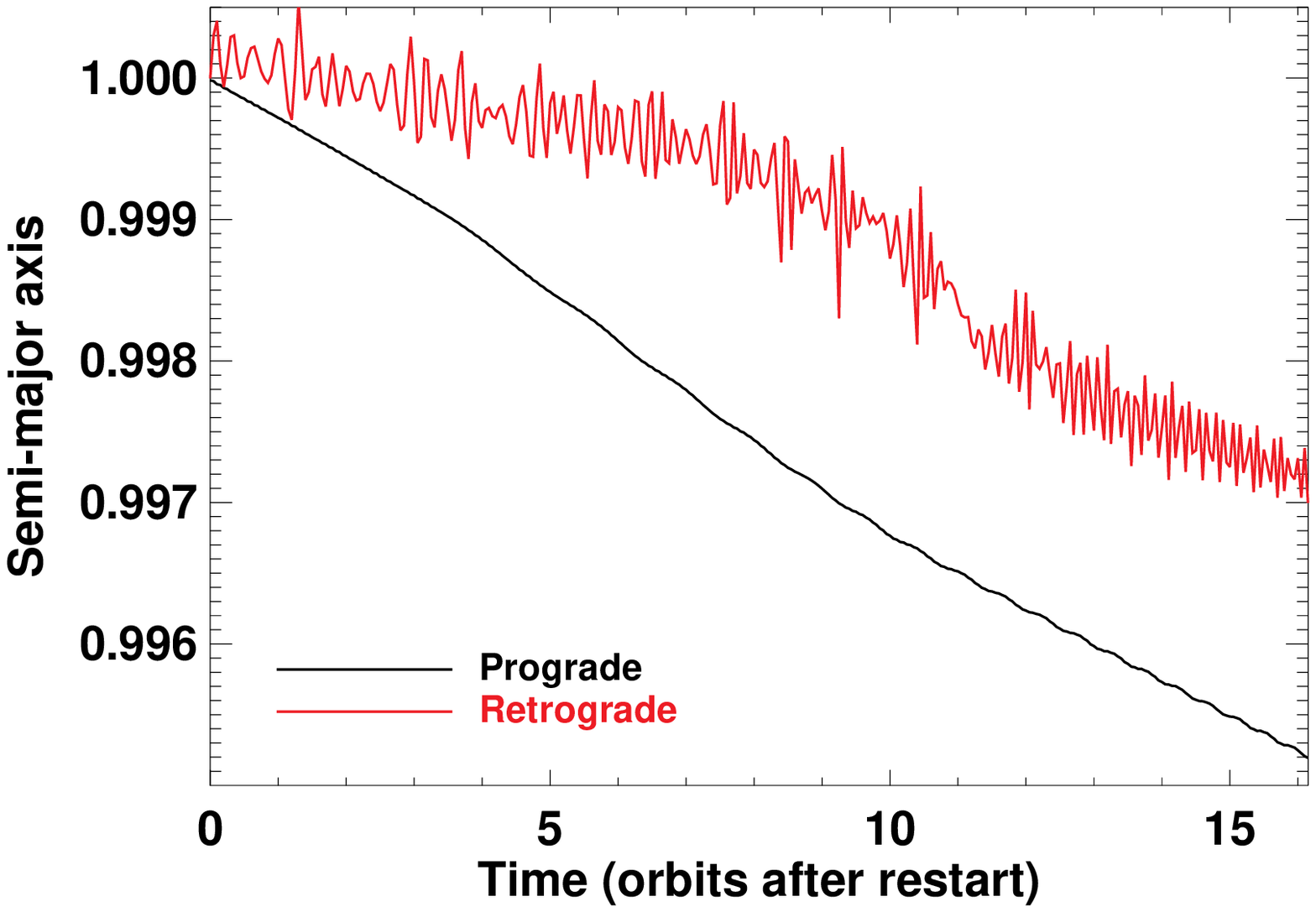}
  \caption{\label{retrograde}Orbital evolution of retrograde and
    prograde binaries. Hardening rates and migration rates are
    compared in the left and right panels, respectively.  The
    horizontal dash-dotted line in the left panel shows the value of
    the stars softening length.}
\end{figure*}
%FFFFFFFFFFFFFFFFF
The calculation results presented in previous sections assume a
prograde binary, with the angular momentum vectors of the disk and of
the binary being aligned. It is possible that this relative
inclination takes very different values when the binary forms in the
disk, or after it formed, through dynamical interactions with other
stars or more massive objects. Another possibility is that the binary
did not form in the disk, and was captured. One expects that the
gaseous disk dampens the relative inclination on a rather short
timescale. Since our calculation results suggest that the binary's
hardening may also occur on short timescales, it is relevant to
investigate the general case of a random disk--binary relative
inclination. We limit ourselves to the case of a retrograde
binary. The binary's center of mass still has a prograde motion around
the black hole.

We performed one simulation with a retrograde binary, using the
procedure and parameters described at the beginning of
\S~\ref{sec:results}. Akin to the prograde case shown in
Figure~\ref{densplot}, the disk's surface density obtained at $500$
orbits with the retrograde binary is displayed in the left panel of
Figure~\ref{densplotR} (note that the color scales differ in both
figures). We observe that the gap is less deep in the retrograde
case. We have measured that the azimuthally-averaged density in the
gap is a factor of $2$ to $3$ larger compared to the prograde
case. The right panel of Figure~\ref{densplotR} shows a close-up of
the disk density around the retrograde binary, whose Hill radius is
drawn as a dashed circle. In this panel, the stars rotate clockwise. A
comparison with the top-right panel of Figure~\ref{densplot} reveals
that the mass enclosed in the Hill radius is about one order of
magnitude larger in the retrograde case. We attribute these
differences to a larger velocity difference in the retrograde case
between the stars and the fluid elements approaching the stars on
horseshoe trajectories. Those fluid elements are therefore less
deflected as they get closer to the stars, and they can enter the
binary's Hill radius more easily.

Similarly as in the prograde case, overdense spiral tails lag the
stars. The tails should therefore extract angular momentum from the
retrograde binary, which we expect to harden with time, just like the
prograde binary. This is confirmed in the left panel of
Figure~\ref{retrograde}, where the time evolution of the binary's
semi-major axis is displayed for both the prograde and the retrograde
binaries. The horizontal dash-dotted curve depicts the stars softening
length. The hardening timescale is a factor of $5$ to $10$ shorter
than in the prograde case, which can be attributed to the relative
difference of the density in the Hill radius between the retrograde
and prograde binaries. The right panel of Figure~\ref{retrograde}
compares the semi-major axis of the binary's center of mass in both
cases. The inward migration of the retrograde binary is much slower on
average, which is most presumably due to a shallower gap profile.

% ========================  
\section{Back to the original problem}
% ========================
\label{sec:original}
%FFFFFFFFFFFFFFFFF
\begin{figure}
  \includegraphics[width=\hsize]{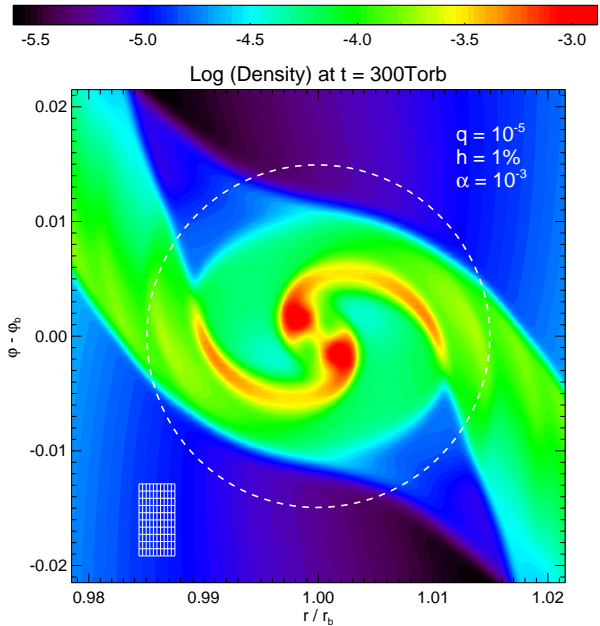}
  \caption{\label{densorig} Surface density contour at $300$ orbits obtained
    with the parameters of the original problem ($q=10^{-5}$, $h=1\%$,
    and $\alpha=10^{-3}$).  The binary's Hill radius is shown as a
    dashed circle, and part of the computational grid is overplotted
    in the bottom-left part of the panel.}
\end{figure}
%FFFFFFFFFFFFFFFFF
%FFFFFFFFFFFFFFFFF
\begin{figure}
  \includegraphics[width=\hsize]{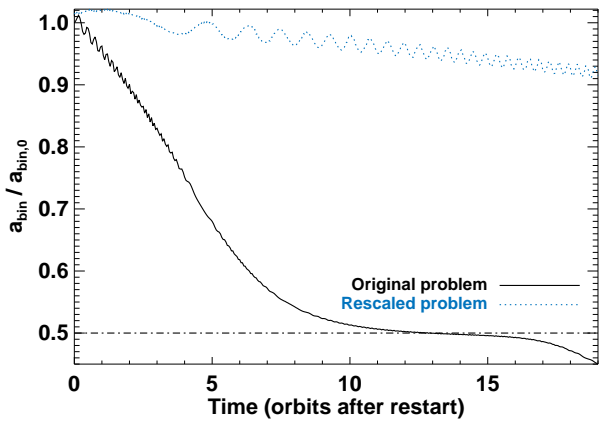}
  \caption{\label{smaorig} Time evolution of the binary's semi-major
    axis obtained with the disk and satellite parameters of the
    original problem (solid curve), and of the rescaled problem with
    same initial density (dashed curve). For comparison purposes, the
    ratio $a_{\rm bin} / a_{\rm bin,0}$ is displayed in y-axis. The
    time in x-axis is in units of the orbital period of the binary's
    center of mass around the central black hole. The horizontal
    dash-dotted line depicts the ratio $\varepsilon / a_{\rm bin,0}$,
    which is identical in both simulations.}
\end{figure}
%FFFFFFFFFFFFFFFFF
Our original problem is to investigate the tidal interaction of a
massive ($q=10^{-5}$) binary star and its natal thin ($h\sim 1\%$)
gaseous disk. We have justified in \S~\ref{sec:rescaledmodel} that,
for a single-star satellite, one can instead tackle a less
computationally demanding rescaled problem, where the properties of
the disk--satellite interaction (gap properties, amount of gas in the
satellite's Hill radius, value of the specific torque on the
satellite) remain essentially unchanged. The calculation results in
\S~\ref{sec:results} have been obtained with the parameters of the
rescaled problem: $q=10^{-3}$, $h = 5\%$ and $\alpha = 4 \times
10^{-3}$.  We have shown that, in addition to migrating inward, the
binary system hardens, due to the net drag force exerted by each
companion's wake inside the binary's Hill radius. Our results indicate
that the hardening timescale is inversely proportional to the
unperturbed density in the binary's Hill radius, and that it is
shorter than the migration timescale by typically one to two orders of
magnitude.

We have shown in \S~\ref{sec:componebin} that the gap properties and
the specific torque obtained with single and binary satellites are in
good agreement. This agreement implies that the steady-state migration
rate of a binary satellite should be similar in the original and
rescaled problems, as in the case of a single satellite detailed in
\S~\ref{sec:rescaledmodel}. However, it is not clear whether the
hardening rate remains unchanged when applying the rescaling
method. To check this, we performed a simulation with the parameters
of the original problem. The unperturbed surface density at the
binary's location is $10^{-4}$, as in the simulation of
\S~\ref{sec:migration}. This sets the initial Toomre parameter at the
binary's location to $\approx 32$. For comparison purposes, we kept
the same ratios $a_{\rm bin} / R_{\rm H}$ and $a_{\rm bin} /
\varepsilon$, which leads to $a_{\rm bin} \approx 4\times 10^{-3} a$
and $\varepsilon \approx 2\times 10^{-3} a$. The satellite's Hill
radius is resolved by about $50$ cells along the radial direction, and
$24$ along the azimuthal direction (to be compared with $50$ and $33$,
respectively, in the simulations of \S~\ref{sec:results}). A close-up
of the gas surface density around the binary star is displayed at
$300$ orbits in Figure~\ref{densorig}. It is very similar to the
density obtained at $500$ orbits in the rescaled problem (top-right
panel of Figure~\ref{densplot}). This similitude is due to (i) the use
of (almost) the same dimensionless parameters in the gap-opening
criterion, and to (ii) the fact the density in the satellite's Hill
radius reaches a faster steady-state than the density in the gap. We
notice that the density at each perturber's wake is a factor of
$\approx 2$ larger in the original problem.

The time evolution of the binary's semi-major axis obtained with
previous simulation restarted at $300$ orbits is depicted in
Figure~\ref{smaorig} (solid curve). To compare with the result of the
rescaled problem (dashed curve), the ratio $a_{\rm bin} / a_{\rm
  bin,0}$ is displayed in y-axis. The binary's semi-major axis in the
original problem decreases by a factor of two in about $8$ to $10$
orbits, or in about $130$ orbits of the stars around their center of
mass. This corresponds to $\sim 1.5 \times 10^4$ yrs at $0.1$ pc in
the Galactic center case. The slowing down of the binary's hardening
rate after $\approx 8$ orbits is most presumably due to the fact that
$a_{\rm bin}$ becomes comparable to the softening length of the stars
potential. The hardening timescale in the original problem is shorter
than that of the equivalent rescaled problem by a factor of $8$ to
$10$. This timescale difference can be qualitatively interpreted as
due to different mass ratios of the binary and of the gas in the
binary's Hill radius. Between the original and rescaled problems, the
binary's mass is increased by a factor of $100$, whereas the mass
inside the binary's Hill radius in increased by a factor of $\sim 20$
(ratio of the Hill radii squared). This simple argument suggests a
hardening timescale $\sim 5$ times larger in the rescaled problem,
which is in correct agreement with our findings.

Some further insight can be gained by using the analytic prediction of
\cite{s10}. Through a linear analysis of the Euler equations, he
derived the timescale for the orbital decay of a low-mass binary
embedded in a three-dimensional static medium of gas. Applying his
equation 56-a to our disk model (writing the sound speed $c_{\rm s} =
H \Omega$, the 3D unperturbed gas density $\rho_0 \approx \Sigma_0 /
2H$, noticing from our simulations that a typical density in the
binary's Hill radius is the unperturbed density at the binary's
initial location), the extrapolated hardening timescale $\tau_{\rm
  hard}$ may be estimated as
\begin{equation}
  \tau_{\rm hard} \sim 0.6 \left( \frac{Q}{h} \right) 
  \left( \frac{q}{h^3} \right)^{-2} 
  \left( \frac{a_{\rm bin}}{a} \right) T_{\rm orb},
  \label{stahler}
\end{equation}
where $T_{\rm orb}$ is the orbital period of the binary's center of
mass around the central black hole. Since our original and rescaled
simulations have the same $Q/h$ (that is, the same unperturbed density
at the binary's location), Eq.~(\ref{stahler}) predicts that the
hardening timescale in the original problem should be a factor of $8$
shorter than in the rescaled problem, which is in good agreement with
our results of simulations. For convenience, Eq.~(\ref{stahler}) can
be recast as
\begin{eqnarray}
  \left.  \tau_{\rm hard}
  \right. & \sim & 900\;{\rm yr}
  \times
  \left( \frac{T}{2000\;{\rm K}} \right)^{3}
  \left( \frac{\Sigma}{10^2\;{\rm g\;cm}^{-2}} \right)^{-1}
  \nonumber\\
  &\times&
  \left( \frac{M_{\rm smbh}}{3\times10^6\;M_{\odot}} \right)^{-1/2}
  \left( \frac{M_{\rm bin}}{2\;M_{\odot}} \right)^{-2}
  \nonumber\\
  &\times&
  \left( \frac{a}{\rm 0.1\;pc} \right)^{3/2}
  \left( \frac{a_{\rm bin}}{\rm 50\;AU} \right),
  \label{stahler2}
\end{eqnarray}
where $M_{\rm bin}$ denotes the binary's mass, and where $T$ and
$\Sigma$ are the unperturbed gas temperature and surface density at
the binary's location. In Eq.~(\ref{stahler2}), we have assumed a mean
molecular weight $\mu=2.4$, and an adiabatic index $\gamma=5/3$.

With the parameters of the original problem, Eq.~(\ref{stahler})
predicts a hardening timescale $\lesssim 0.1$ orbit, that is about two
orders of magnitude shorter than in our simulation. We comment that,
while Eq.~(\ref{stahler}) may capture the right functional dependence
of the hardening timescale with the disk and satellite parameters in
our problem, a quantitative comparison with our results of simulations
seems inappropriate, since the key assumption in \cite{s10}'s analysis
(each perturber's wake induces a small perturbation of the background
gas density, which can be described with a linear analysis) is clearly
not satisfied in our simulations (see e.g., Figure~\ref{densorig}).
Another source of discrepancy may also be triggered by the
time-varying velocity difference in our simulations between the binary
and the gas inside the Hill radius.

% ========================  
\section{Discussion}
% ========================
\label{sec:discu}

% --------------------------------------  
\subsection{Motivation}
% --------------------------------------
The primary astronomical motivation of the investigation presented
here is to explore the origin of the ``S-stars'', a population of
relatively massive stars in the inner $0.05$ pc of the Galactic
center. The S-stars could be related to those in coplanar orbits just
outside this inner region. It is natural to assume that these
disk-stars may have either coherently formed in \citep{levin07} or
rapidly accreted gas \citep{alp93} from a common gaseous disk.

Most of the S-stars have large eccentricities and random inclinations
\citep[e.g.,][]{lu09}. In contrast, the nearby disk-stars have smaller
eccentricities and their velocity dispersion is an order of magnitude
smaller than the local Keplerian speed. One potential explanation for
their large eccentricities and random inclinations is that the S-stars
originated from the tidal disruption of binary systems that were
scattered to the proximity of the central supermassive black hole on
nearly parabolic orbits \citep{gq03}. This scenario would also account
for the observed hypervelocity stars in the Galactic halo, provided
some of the scattered binaries have very close separation
\citep{Hills88, YT03, gualandris05, lzy10a, lzy10b}, typically within 
a few solar radii.

In this paper, we consider a unified scenario for the young, massive
stars near the Galactic center (both disk- and S-stars) and
high-velocity stars in the Galactic halo. We assume their progenitors
were binary stars that formed at large distance (a fraction of a pc)
from the SMBH, in a gaseous disk that we expect existed in the
Galactic center in the past. In principle, similar to the solar
neighborhood, binary systems could be as common as single stars and
have a logarithmic period distribution \citep{DM91}. Around the
Galactic center, however, tidal interaction between binary systems and
their natal disk induces angular-momentum exchange between them. The
main outstanding issue is whether these binaries' center of mass would
migrate (presumably inward) extensively before they harden
substantially.

% --------------------------------------  
\subsection{Model}
% --------------------------------------
We have investigated the tidal interaction of a binary star and its
natal gaseous disk by means of two-dimensional hydrodynamical
simulations. For illustration purpose, we have considered a fiducial
binary system, comprising two equal-mass ($15\;M_{\odot}$) stars
embedded in a thin ($h = 1\%$), moderately turbulent ($\alpha =
10^{-3}$) gaseous disk around a $3\times 10^{6}\;M_{\odot}$
SMBH. These initial and boundary conditions are such that the binary
opens a deep gap around its orbit, and is thus subject to type II
migration. We have shown in \S~\ref{sec:rescaledmodel} that this
gap-opening process has helped us transform our original problem into
an equivalent, less computationally demanding rescaled problem. We
have presented results of calculations of the rescaled problem in
\S~\ref{sec:results}, and of the original problem in
\S~\ref{sec:original}.

We have made a number of simplifying assumptions in the models
presented here (see \S~\ref{sec:model}). We have neglected the process
of gas accretion onto the stars, assuming the latter already reached
their final mass at the beginning of the simulations.  We have also
assumed throughout this paper that the binary stars formed in a
circular orbit, while the fragmentation of an eccentric gaseous disk
would put them in initially eccentric orbits \citep[e.g.,][]{ncs07,
  Alexander08ecc}. While we expect many stars to form at the same time
in a fragmenting gas disk, we have considered the evolution of a
single isolated satellite (one binary star). The presence of many
embedded stars, each propagating density waves, and opening a gap for
the most massive ones, would dramatically impact the gas disk
structure. The properties of disk--satellite interactions, and the
resulting migration timescale, could be very different from those of a
single isolated satellite. The interaction between different
satellites could excite their orbital eccentricity and alter their
migration rate, leading for example to captures into mean-motion
resonance or scattering events \citep[e.g.,][]{CN08, MBS10}. Also,
encounters with other single stars may help to further harden binary
stars, especially at short separations from the supermassive black
hole \citep[e.g.,][]{peretshv09}. We defer this interesting regime for
future study.

The properties of the gaseous disk after star formation are largely
uncertain and poorly constrained. We have grossly simplified the
evolution of the underlying gaseous disk, in particular by neglecting
the depletion of the gaseous disk, and the increase of its temperature
under star formation. Local heating of the gas disk by the intrinsic
luminosity of the embedded stars could modify the mass flow inside the
binary's Hill radius, and therefore the binary's hardening rate. The
progressive depletion of the gas disk also implies a finite timescale
over which disk--satellite interactions are effective. In the context
of Sgr A*, we mention the work of \cite{ncs07}, who performed
hydrodynamical simulations of star formation in a gaseous disk around
the central SMBH, including a crude treatment of thermal accretion
feedback (a fraction of the star accretion luminosity is converted
into disk heating). Their simulations indicate that the disk's mass
may be reduced by a factor of two in at most $10^5$ yrs (about $60$
orbital periods at $0.1$ pc). It is much shorter than the timescale
for gas accretion onto Sgr A* through turbulent transport of angular
momentum \citep{nc05}. It is also much shorter than the expected
timescales for type I and type II migration, but not necessarily short
compared to the timescale for type III migration (see
\S~\ref{sec:single}). Still, this potentially short lifetime of the
gas disk is a challenge for migration driven by disk torques. However,
it may not be a stringent constraint on the hardening of massive
binary stars driven by the gas disk, which we find to occur on a
timescale comparable to, or shorter than the aforementioned disk
lifetime, as is recalled in \S~\ref{ref:summary}.

% --------------------------------------  
\subsection{Summary of our results}
% --------------------------------------
\label{ref:summary}
Our main result is that the binary gets harder as it migrates toward
the central black hole. The hardening timescale is much shorter than
the migration timescale, which is very similar to that of a single
satellite of the same mass. The binary's hardening rate is essentially
determined by the gas density inside the binary's Hill radius. In that
region, overdense spiral tails lag the stars, exerting a torque on
them and hardening their orbit. The gas density at the tails location,
and therefore the hardening rate, scale with the viscosity and with
the unperturbed surface density of gas at the binary's location ({\it
  i.e.} prior to gap formation). Assuming that at this location, the
$\alpha$ viscous parameter is $\sim 10^{-3}$, and the initial Toomre
parameter (before gap formation) is $\approx 32$, we find that the
separation between the stars is reduced by a factor of $2$ in about
$10$ orbits of the binary around the central black hole (see
\S~\ref{sec:original}). Scaled to the Galactic center case, that is in
only $\sim 2\times 10^4$ yrs at $0.1$ pc, which is several orders of
magnitude shorter than the expected migration timescale. A larger
initial surface density would yield an even shorter hardening
timescale. In the rest of this section, we discuss the evolution of
low-mass binary stars (namely, what could have happened prior to our
initial condition), and the subsequent evolution of massive binary
stars (that is, what could occur after the end of our simulations).

% --------------------------------------  
\subsection{Hardening of low-mass binary stars}
% --------------------------------------
It is likely that disk--satellite interactions significantly impact
the binary's semi-major axis from the early stages of the binary's
formation, well before the binary has attained its asymptotic mass and
potentially opened a gap around its orbit. In that case, the wake of
each star would induce a low-amplitude density perturbation in the
background gas surrounding the stars. This situation would resemble
the one investigated by \cite{s10}. The hardening timescale
extrapolated from his analytic study to our disk model is given at
Eqs.~(\ref{stahler}) and~(\ref{stahler2}). Assuming the unperturbed
gas disk has $Q \sim 32$, $h\sim 1\%$ (that is, $\Sigma \sim 6$ g
cm$^{-2}$ and $T\sim 2000$ K at the binary's location), and taking a
low-mass binary ($1\;M_{\odot}$ stars, so that $q/h^3 \lesssim 1$ and
the linearization assumption should be valid) with $a_{\rm bin} \sim
0.3R_{\rm H}$ (as in our simulations; here it corresponds to $a_{\rm
  bin} \sim 40$ AU), Eqs.~(\ref{stahler}) or~(\ref{stahler2}) give
$\tau_{\rm hard} \approx 1.2\times 10^4$ yr at $a = 0.1$ pc. This
timescale is similar to the one obtained in our 2D simulations, but
note the strong dependence of $\tau_{\rm hard}$ with both $q$ and
$h$. The dynamical evolution of a low-mass binary subject to hardening
during its type I migration deserves follow-up investigation with
self-consistent high-resolution 3D simulations.

% --------------------------------------  
\subsection{Subsequent binary evolution}
% --------------------------------------
The results in \S~\ref{sec:results} and~\ref{sec:original} provide the
initial hardening rate for massive (gap-opening) binary stars with
separation close to their Hill radius. We underscore that our
simulations cannot, and do not address the final outcome of the
binary's hardening, {\it i.e.} whether binary stars will ultimately
coalesce, form a very hard binary, or become disrupted. We can only
extrapolate some possible implications of the binary's fast hardening
in our model.

% - - - - - - - - - - - - - - - - - - - - - -
\subsubsection{Opening of a cavity}
% - - - - - - - - - - - - - - - - - - - - - -
\label{sec:cavity}
As a massive binary star shrinks, residual gas is preserved in a
viscous disk between its orbit and its Hill radius. Tidal interaction
between the binary and the disk around it leads to angular momentum
transfer from the stars' orbits to the inner disk region. Similarly,
angular momentum is removed from the outer region of the disk by the
SMBH's tidal torque. The binary's hardening rate is determined by an
equilibrium (outward) flux of angular momentum. Our results suggest
this rate may be maintained at a fairly large value at least until the
binary separation is less than about one sixth of its Hill radius. Our
results also suggest that the disk interior to the binary stars' orbit
should be cleared by their tidal torque such that this region is
expected to evolve into a cavity. In that case, the inside of the
binary's Hill radius would comprise the binary star, the cavity where
the binary now evolves in, and the surrounding gas disk referred to as
the circumbinary disk (the cavity should not be confused with the gap
that the satellite opens or has opened in the global disk structure
around the SMBH).  It remains however to be clarified how the
timescale associated with gas removal from inside the binary's orbit
compares with the binary's shrinking timescale before a cavity is
opened. In the next paper of this series, we will present
high-resolution hydrodynamical simulations to make this
comparison. \emph{If} the binary manages to maintain a cavity, the
tidal interaction between the binary and the circumbinary disk would
keep on hardening the binary, at slower rates than the initial pace we
found here. Otherwise, the binary could keep on hardening until it
coalesces. It is thus possible that some of the massive stars near the
Galactic center result from coalescences of binary-star systems, which
might contribute to account for the top-heavy initial mass function
observed in the two stellar disk structures in the Galactic center
\citep{bartko09l}.  However, such stars would still be embedded in a
gaseous disk around the SMBH, and therefore in nearly-circular
orbits. Resonant relaxation \citep[e.g.,][]{rt96, perets09ecc}, the
presence of short-lived massive stars or of intermediate mass black
holes \citep[e.g.,][]{yll07}, or an initially-eccentric gaseous disk
\citep[e.g.,][]{cuadra08} could explain their observed current
eccentricities.

The hardening process with a cavity inside the binary's orbit is very
similar to type II migration: the tidal torque of the binary competes
with the viscous torque from the circumbinary disk \citep[see
e.g.,][]{Pringle91, arty91, sc95, ipp99}. This problem shares a number
of analogies with the hardening of binary black-holes surrounded by a
circumbinary disk \citep[e.g.,] [where the transport of angular
momentum is due to the disk self-gravity] {cuadra09}. We take the
separation $r_{\rm o}$ between the binary's center of mass, and the
location in the circumbinary disk where most of the binary's angular
momentum is deposited (that is, near the inner edge of the
circumbinary disk), equal to $a_{\rm bin}$. Further assuming that the
gas density $\Sigma$ at that location is still as large as $\sim
10^2\;{\rm g\;cm}^{-2}$, the ratio $4\pi \Sigma a^2_{\rm bin} / M_{\rm
  bin} \sim 5\times 10^{-2} \ll 1$ for a $M_{\rm bin} = 30\;M_{\odot}$
binary with separation $a_{\rm bin} \sim 100\;{\rm AU}$.  The
asymptotic hardening timescale $\tau_{\rm hard, as}$ of the binary
thus corresponds to the timescale for satellite-dominated type II
migration (see \S~\ref{sec:typeII}). Using Eq.~(\ref{tauIIb}), we have
\begin{equation}
  \tau_{\rm hard, as} \sim \frac{M_{\rm bin}}{6\pi\nu\Sigma},
\label{ipp1}
\end{equation}
where $\Sigma$ and $\nu$ denote the surface density and kinematic
viscosity at $r_{\rm o}$. Further writing $\nu = \alpha c_{\rm s}^2
\Omega^{-1}$, with $c_{\rm s}$ and $\Omega$ the sound speed and
angular velocity of the circumbinary disk, we obtain
\begin{eqnarray}
  \left.  \tau_{\rm hard, as}
  \right. & \sim & 9\;{\rm Myr}
  \times \left( \frac{\alpha}{10^{-3}} \right)^{-1}
  \left( \frac{T}{2000\;{\rm K}} \right)^{-1}
  \left( \frac{M_{\rm bin}}{30\;M_{\odot}} \right)^{3/2}
  \nonumber\\
  &\times&
  \left( \frac{\Sigma}{10^2\;{\rm g\;cm}^{-2}} \right)^{-1}
  \left( \frac{a_{\rm bin}}{\rm 100\;AU} \right)^{-3/2},
  \label{ipp2}
\end{eqnarray}
where we have assumed a mean molecular weight $\mu=2.4$, and an
adiabatic index $\gamma=5/3$. The unknown values of $\alpha$, $T$ and
$\Sigma$ near the inner edge of the circumbinary disk make the value
of $\tau_{\rm hard, as}$ largely uncertain.  In order to compare with
the hardening timescale inferred from our simulations, we arbitrarily
assume that $\alpha$, $\Sigma$ and $T$ take similar values as in our
physical model in \S~\ref{sec:model}, which however describes the
properties of the star-forming gas disk around the SMBH. This choice
is also motivated by the simplifying assumption that the circumbinary
disk is in a quasi-steady state. The quantity $\nu\Sigma$ in
Eq.~(\ref{ipp1}) then becomes independent of the separation from the
binary's center of mass, and it may be evaluated near the outer edge
of the circumbinary disk (where $\alpha$, $c_{\rm s}$ and $\Omega$
should not dramatically differ from their value in the star-forming
disk about the central SMBH, at the binary's location).  Thus taking
$T \sim 2000$ K, $\Sigma \sim 10\;{\rm g\;cm}^{-2}$, $\alpha \sim
10^{-3}$, $M_{\rm bin} = 30M_{\odot}$, and $a_{\rm bin} \sim 100$ AU,
we get $\tau_{\rm hard, as} \approx 90$ Myr. This timescale is, as
expected, much longer than the hardening timescale with no cavity
inside of the binary's orbit. Even with considering a more
conventional larger $\alpha$, smaller values of $\Sigma$ due to the
progressive disk depletion should maintain $\tau_{\rm hard, as}$ to a
value that presumably exceeds a few Myr. This is comparable to, if not
larger than the lifetime of the most massive stars near the Galactic
center, indicating that stellar evolution should be taken into account
(see \S~\ref{sec:evol}).

We finally point out that there is no indication in Eq.~(\ref{ipp2})
as to why the binary's shrinkage should stop. We mention the
possibility that the gas in the circumbinary disk may be progressively
stripped due to stellar winds, or to multiple close encounters with
other stars, as was observed by \cite{ncs07}. If the shrinkage is
sufficiently efficient to decrease the binary's separation to a few
stellar radii, mass transfer due to Roche lobe overflow will control
the binary's subsequent evolution. If the donor star expands faster
than its Roche lobe, or shrinks less rapidly than its Roche lobe for a
prolonged time, mass transfer will be unstable, which may lead to a
common-envelope phase and, perhaps, a merger.

% - - - - - - - - - - - - - - - - - - - - - -
\subsubsection{Including stellar evolution: the explosive disruption scenario}
% - - - - - - - - - - - - - - - - - - - - - -
\label{sec:evol}
%FFFFFFFFFFFFFFFFF
\begin{figure*}
  \includegraphics[width=0.33\hsize]{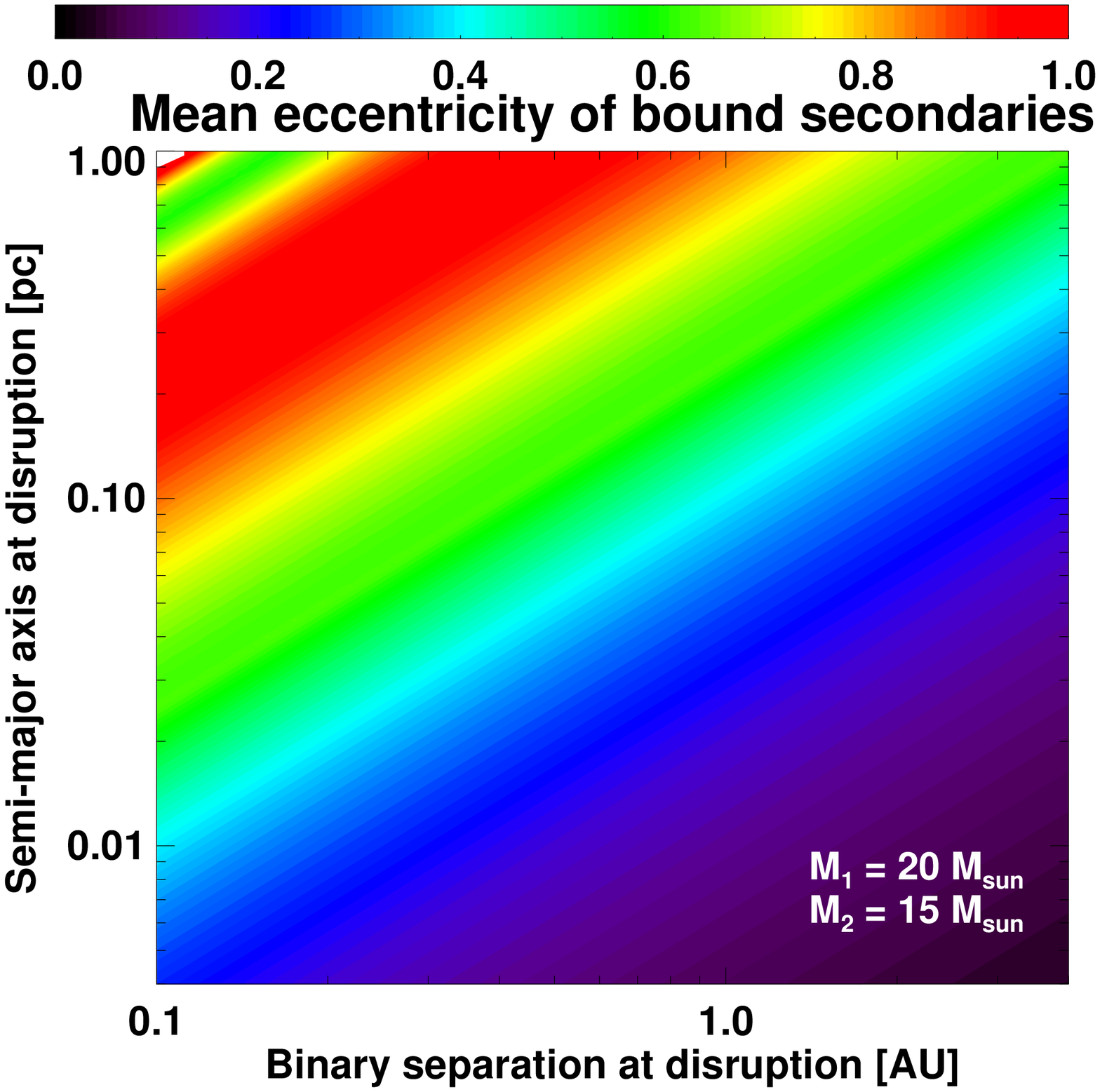}
  \includegraphics[width=0.33\hsize]{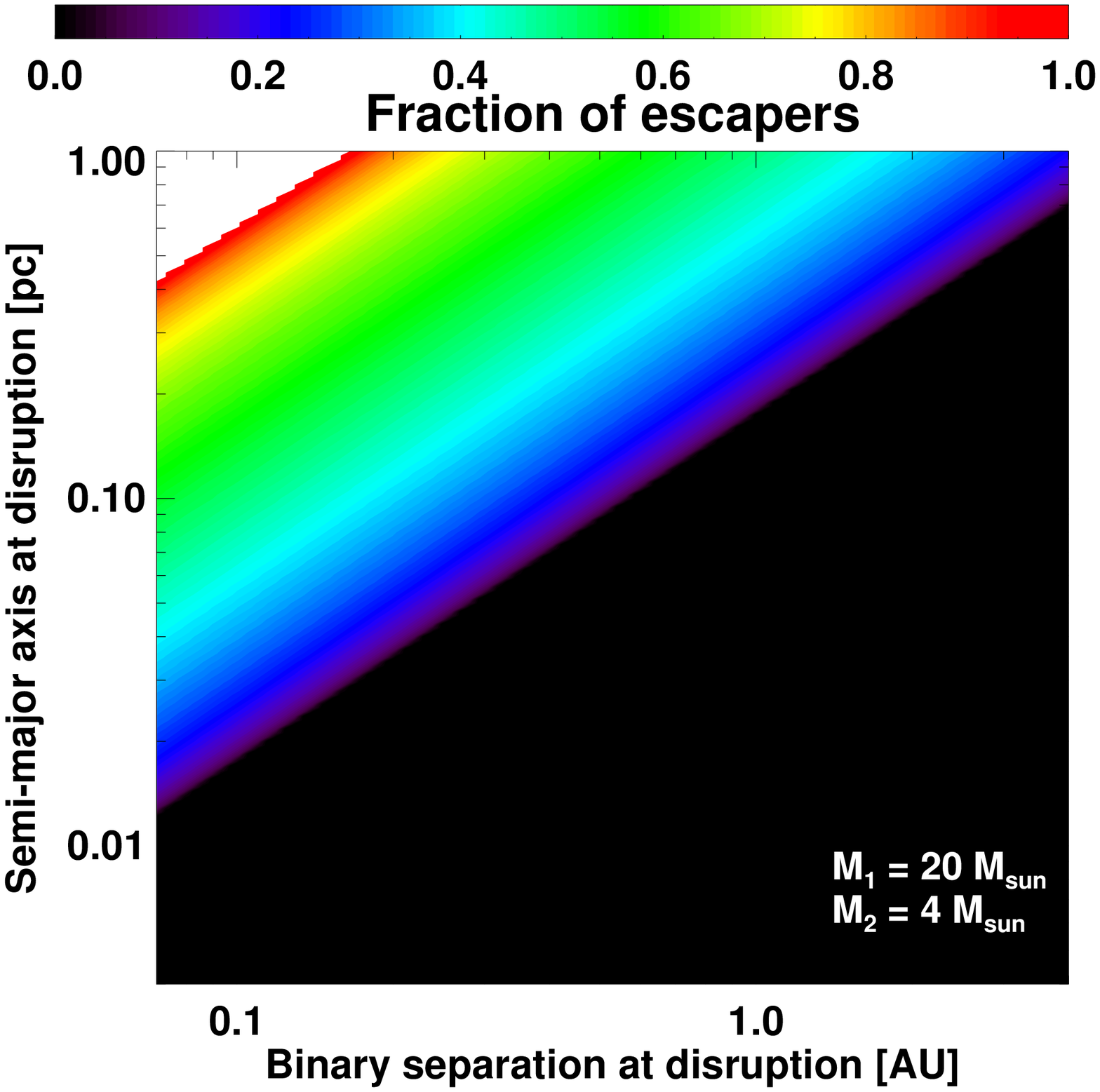}
  \includegraphics[width=0.33\hsize]{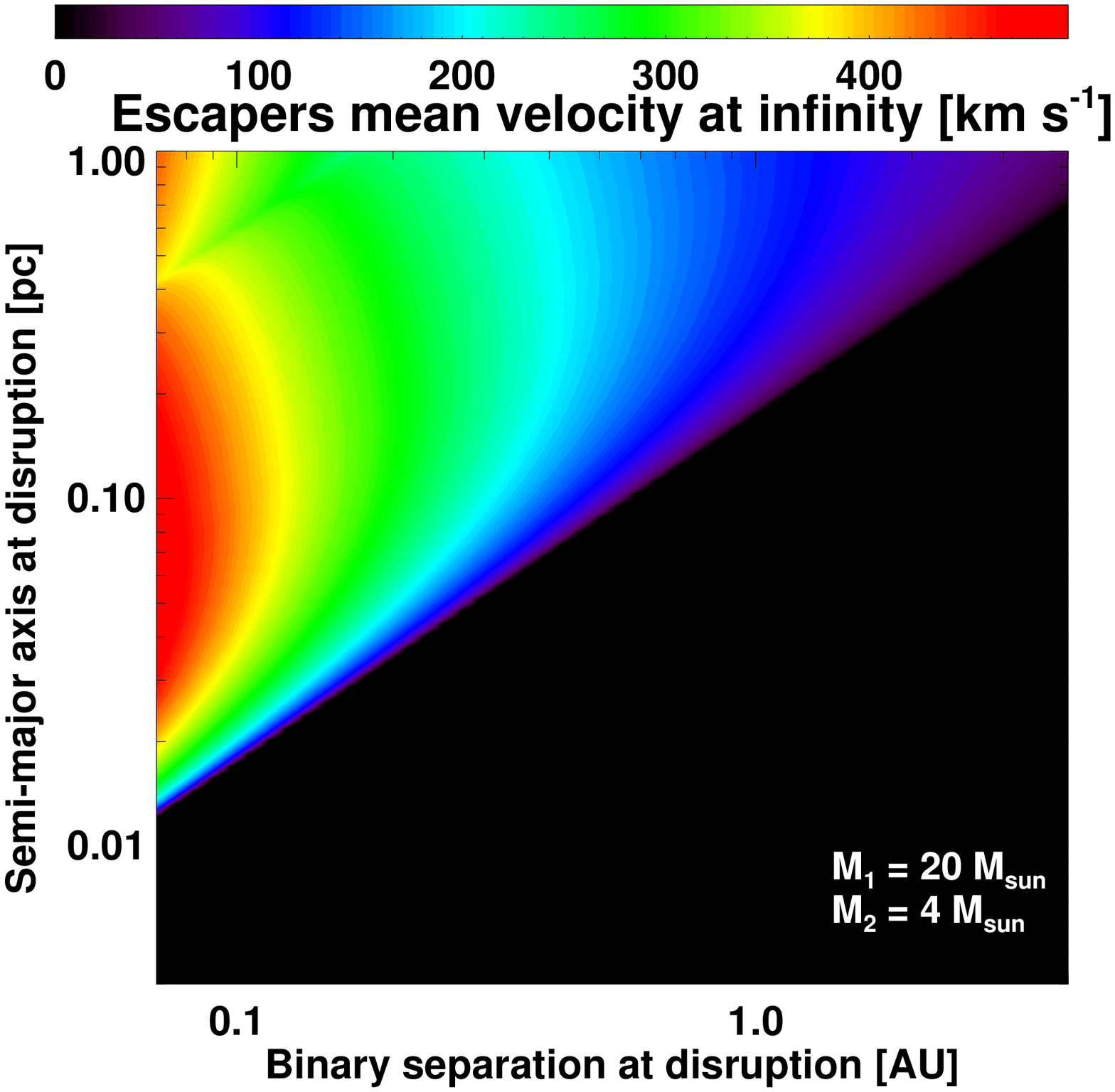}
  \caption{\label{disruption}Outcome of the supernova disruption of a
    binary star orbiting a $M_{\rm smbh} = 3.5\times 10^6 M_{\odot}$
    black hole. Before disruption, the primary's mass is $M_1 = 20
    M_{\odot}$, and the secondary's mass is either $M_2 = 15
    M_{\odot}$ (left panel), or $M_2 = 4 M_{\odot}$ (middle and right
    panels). We assume that the primary loses enough mass so that the
    binary becomes disrupted. Before disruption, the internal and
    orbital angular momentum vectors of the binary are taken aligned.
    We assess in this figure the properties of the two-body orbit of
    the secondary with the black hole (see text for details). In all
    contour plots, the binary's internal separation $a_{\rm bin}$ is
    depicted in x-axis, and the separation $a$ between the black hole
    and the binary's center of mass is in y-axis. The left panel shows
    the mean eccentricity of the bound secondaries. The middle panel
    displays the fraction of escapers, namely the secondaries that
    become unbound to the black hole. The escapers' mean velocity at
    infinity is shown in the right panel.}
\end{figure*}
%FFFFFFFFFFFFFFFFF
Including stellar evolution, and the likely possibility of
unequal-mass binaries, we now consider the outcome of binaries that
have reached small enough separations. If the more massive (primary)
star evolves into a red giant with a radius comparable to the binary
separation, both stars undergo common envelope evolution. The engulfed
binary hardens again due to the drag force exerted by the gas on each
star. The energy lost by the binary system leads to the expansion of
the gas envelope. This is reminiscent of the opening of a cavity by a
binary embedded in a gaseous disk, discussed in
\S~\ref{sec:cavity}. The successful ejection of the common envelope,
and therefore the binary's survival, depend on the binary separation
before the common envelope stage, and on the primary's mass \citep[see
e.g., the review by][]{Taam00}. If the binary survives the common
envelope evolution ({\it i.e.} if it does not coalesce), it then
comprises an Helium-rich star and an OB-star companion.

Many stars near the Galactic center lose mass through Wolf-Rayet
winds. With a substantial mass loss, the primary may evolve into
Wolf-Rayet or blue supergiant stars. Mass loss typically results in
increasing the binary separation, but in our scenario this trend is
offset by the disk presence. In this case, the binary separation can
be reduced to several solar radii, and may avoid another common
envelope evolution if the primary explodes as a supernova.  The
outcome of the primary's explosion depends on the fraction of the
primary's mass that is ejected. If they explode, typical blue
supergiants (with a mass $\sim 10-20\;M{_\odot}$) may lose most of
their residual envelope during their supernova explosion. If the mass
lost to the supernova ejecta is more than half of the binary's mass
before explosion, the compact binary would become unbound
\citep{Hills83}, and the secondary could attain a recoil velocity that
is a significant fraction of the primary's surface escape velocity.

What velocities can we expect in the supernova disruption scenario?
The simplest approach is to look at the properties of the two-body
orbit of the secondary about the central black hole. Before
disruption, we assume that the binary's center of mass is on a
circular orbit with semi-major axis $a$, and the binary's internal
separation is $a_{\rm bin}$ (we neglect the binary's internal
eccentricity $e_{\rm bin}$ for simplicity).  We also assume that the
primary's supernova has no impact on the secondary: the mass and
velocity of the secondary before and after disruption are taken the
same. We will come back to this assumption below. In the frame
centered on the black hole, the velocity of the secondary is the sum
of the velocity of the binary's center of mass about the black hole
(${\bf v}_{\rm smbh}$), and of the velocity of the secondary about the
binary's center of mass (${\bf v}_{\rm bin}$). The angle $\theta$
between both velocity  vectors is a random variable distributed
uniformly between $0$ and $2\pi$, since in our model both velocities
are coplanar. Their amplitude is given by
\begin{equation}
  v_{\rm smbh} \sim 390\;{\rm km\;s^{-1}} \times \left( \frac{M_{\rm smbh}}{3.5\times 10^6 \;M_{\odot}} \right)^{1/2} \left( \frac{a}{0.1\;{\rm pc}} \right)^{-1/2},
\label{disru1}
\end{equation}
and 
\begin{equation}
  v_{\rm bin} \sim 133\;{\rm km\;s^{-1}} \times 
  \left( \frac{M_1}{M_{\rm bin}} \right)
  \left( \frac{M_{\rm bin}}{20\;M_{\odot}} \right)^{1/2}\left( \frac{a_{\rm bin}}{1\;{\rm AU}} \right)^{-1/2},
\label{disru2}
\end{equation}
respectively. In Eq.~(\ref{disru2}), $M_1$ and $M_2$ denote the mass
of the primary and of the secondary, and $M_{\rm bin} = M_1 + M_2$ is
the binary's mass. In the following, we take $M_{\rm smbh} = 3.5\times
10^6\;M_{\odot}$, and $M_1 = 20\;M_{\odot}$.

We first consider the case where $M_2 = 15\;M_{\odot}$, a mass similar
to that of the S-stars. We vary $a$ from $4\times 10^{-3}$ pc to $1$
pc, and $a_{\rm bin}$ from $0.1$ AU to $4$ AU (both with a logarithmic
spacing).  The minimum value of $a_{\rm bin}$ is approximately the sum
of the two stars radii. Its maximum value is set as about half the
smallest Hill radius of our binaries sample. This choice ensures that,
before the primary goes supernova, all binaries are bound with modest
internal eccentricity.  The left panel of Figure~\ref{disruption}
shows a contour plot of the mean eccentricity of the secondaries that
remain bound to the central object, as a function of $a_{\rm bin}$
(x-axis) and $a$ (y-axis). The average is done over the different
values of the angle $\theta$. Large eccentricities, compatible with
those of the S-stars, can be obtained with the supernova disruption
scenario. We point out that, for sufficiently large $a$ (at a given
$a_{\rm bin}$), the mean eccentricity of the bound secondaries
decreases, with a local minimum of $\sim 0.6$, and increases
again. The presence of this local minimum is due to secondaries put on
retrograde, close to circular orbits about the black hole. It is
straightforward to show that the minimum's location satisfies $v_{\rm
  bin} = 2 v_{\rm smbh}$, which can be recast as $a \sim 0.6\;{\rm
  pc}\times (a_{\rm bin}/0.1\;{\rm AU})$ for our set of parameters.

We next consider the case where $M_2 = 4\;M_{\odot}$, a mass similar
to that of the hypervelocity stars, and we now vary $a_{\rm bin}$ from
$0.07$ AU to $4$ AU.  The middle panel of Figure~\ref{disruption}
shows the fraction of escapers, these secondaries that are energetic
enough to become unbound to the black hole. The fraction of escapers
increases with increasing $a$ and decreasing $a_{\rm bin}$, as
expected. For maximally hard binaries, the fraction becomes unity for
$a \gtrsim 0.4$ pc.

The escapers' mean velocity at infinity, measured in the frame
centered on the SMBH, is shown in the right panel of
Figure~\ref{disruption}.  The large velocities that we
obtain\footnote{We have checked that the inclusion of the
  gravitational potential of the Galactic bulge, disk and halo
  \citep[see e.g.,][]{xue08} has a negligible impact on estimating the
  escapers velocity at infinity.}  suggest that the most energetic
escapers could account for some of the Runaway Galactic OB-stars, as
originally proposed by \cite{Blaauw61}, or for some of the
hypervelocity stars in the Galactic halo. The latter have typical
Galactic rest-frame velocities between $400$ and $700\;{\rm
  km\;s}^{-1}$ \citep[e.g.,][]{Brown06b}. Our results indicate that
the ejection of secondaries that may still be going faster than $\sim
400$ km s$^{-1}$ in the Galactic halo is possible.  It however
requires binaries that are nearly maximally hard, and primaries that
are sufficiently massive (the maximum escapers velocity approximately
scales proportional to $\sqrt{M_1}$; a primary's mass before explosion
of $\sim 10M_{\odot}$ would yield a maximum escaper's velocity $\sim
330$ km s$^{-1}$).
  
The suggestion that the hypervelocity stars could originate from the
supernova disruption of binaries close to the SMBH is in apparent
contradiction with the results of \cite{gualandris05}, who performed
population synthesis studies and found that no binary can produce a
hypervelocity star by itself. However, in our scenario, binaries can
be kept at very short internal separations due to the action of the
gaseous disk. In contrast, standard population synthesis studies do
not model the angular momentum loss of binaries in presence of a
background disk, and thus produce few very hard binaries. We stress
that our study mentions the possibility, but does not estimate the
probability to get escapers with velocities comparable to those of the
runaway and hypervelocity stars through supernova-disrupted
binaries. We finally comment that the distribution of the
hypervelocity stars on two thin disks \citep{lzy10a, lzy10b} may be
accounted for by both the tidal and the supernova disruption
scenarios. In both cases, we need compact binaries originating from
two disks near the Galactic center.  While the tidal disruption
mechanism requires binaries on highly elliptical orbits, the supernova
disruption mechanism requires binaries with internal and orbital
angular momentum vectors that are nearly aligned.

\begin{figure}
  \includegraphics[width=\hsize]{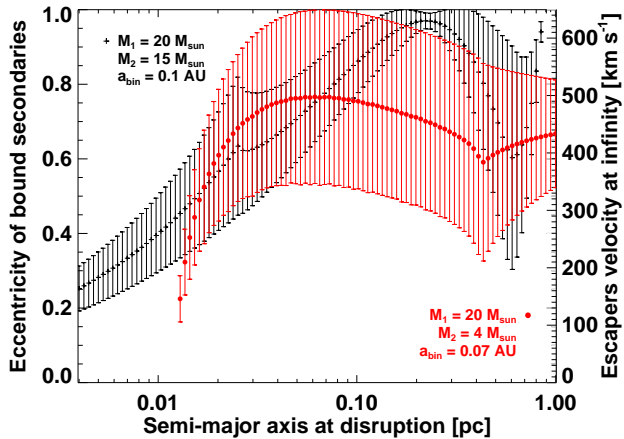}
  \caption{\label{esvel}Eccentricity of the bound secondaries (black
    cross symbols, left axis), and escapers' velocity at infinity (red
    filled circles, right axis) resulting from the supernova
    disruption of nearly maximally hard binaries with aligned internal
    and orbital angular momentum vectors.  Error bars indicate the
    standard deviation of each quantity.}
\end{figure}

Only mean quantities are displayed in Figure~\ref{disruption}. To give
an idea of their standard deviations, we show in Figure~\ref{esvel}
the mean and standard deviation of (i) the eccentricity of the bound
secondaries (black cross symbols, see left axis), and of (ii) the
escapers velocity at infinity (red filled circles, right axis)
resulting from nearly maximally hard supernova-disrupted binaries.
% FFFFFFFFFFFFFFFFF
%FFFFFFFFFFFFFFFFF
%FFFFFFFFFFFFFFFFF
\begin{figure}
  \includegraphics[width=\hsize]{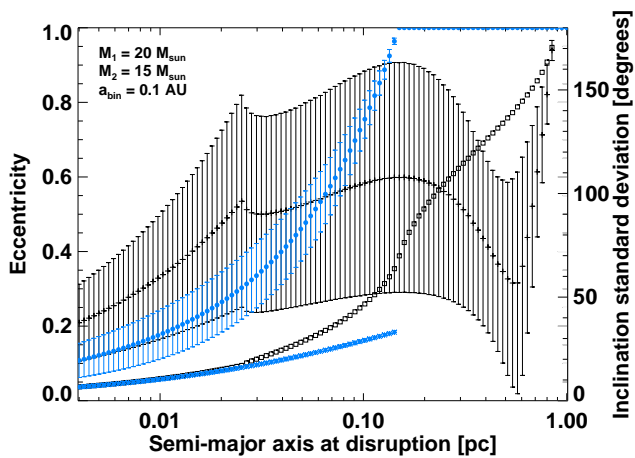}
  \caption{\label{esinc}Eccentricity (mean and standard deviation
    shown as error bars, left axis) and inclination (standard
    deviation, right axis) of the bound secondaries resulting from the
    supernova disruption of nearly maximally hard binaries with
    misaligned internal and orbital angular momentum vectors. Two
    cases are contemplated: one for which the internal angular
    momentum vector of the binary before disruption points towards the
    black hole (black cross and square symbols), and one where it is
    parallel to ${\rm v}_{\rm smbh}$ (blue filled circles and star
    symbols).}
\end{figure}
%FFFFFFFFFFFFFFFFF
  
The results shown in Figures~\ref{disruption} and~\ref{esvel} assume
that the internal and orbital angular momentum vectors of the binary
before disruption are aligned, resulting in a change of the
secondaries' eccentricity, but not of their inclination. Our model
assumes that binaries are formed and stay in the same plane as that of
the gas disk. While a more general scenario, including possible
misalignments and their distribution, is out of this paper's scope, we
briefly underscore the possibility to get  secondaries with large
inclinations through the supernova  disruption scenario. This
statement is illustrated in Figure~\ref{esinc}, where we display the
eccentricity (mean, and standard deviation depicted as error bars),
and standard deviation of the inclination of bound secondaries arising
from the disruption of nearly maximally hard binaries at different
separations from the SMBH ($M_1 = 20 M_{\odot}, M_2 = 15 M_{\odot}$
and $a_{\rm bin} = 0.1$ AU). Only two extreme cases are shown, one
where the internal angular momentum vector of the binary initially
points towards the SMBH (black cross and square symbols), and one
where it is parallel to ${\bf v}_{\rm smbh}$ (blue filled circles and
star symbols). Our results show that it is possible to obtain
secondaries with large inclinations for a broad range of
eccentricities (including large ones) through the supernova disruption
of very hard, circular binaries. In contrast, the tidal disruption of
such binaries can only lead to stars with small inclinations, since
the ratio of the binary's orbital and internal velocities $\sim
(M_{\rm smbh} / M_{\rm bin})^{1/3} \gg 1$. A statistical investigation
aimed at comparing the outcome of the supernova disruption scenario
with the orbital properties of the known S-stars will be presented
elsewhere.

We stress that the above results assume a very simple disruption
model. In particular, we have discarded the impact velocity imparted
by the ejected shell of the primary. Its expression is given by
\cite{tt98}. Still assuming for simplicity that the secondary's mass
is not altered by the supernova's blast wave, the impact velocity
$v_{\rm imp}$ can be estimated as
\begin{equation}
  v_{\rm imp} \sim v_{\rm sh}\times \left( \frac{R_2}{2 a_{\rm bin}} \right)^{2}\left( \frac{M_{\rm sh}}{M_2} \right),
\label{disru3}
\end{equation}
where $M_{\rm sh}$ and $v_{\rm sh}$ denote the mass and ejection speed
of the ejected shell. Taking $v_{\rm sh} \sim 10^4$ km s$^{-1}$
\citep{tt98}, $M_1 = 20M_{\odot}$, $M_2 = 4M_{\odot}$, and $M_{\rm sh}
\sim 12\;M_{\odot}$ (such that the binary loses about half of its
initial mass and becomes unbound), we get $v_{\rm imp} \sim 1.5\;{\rm
  km\;s^{-1}}\,(a_{\rm bin}/1\;{\rm AU})^{-2}$, that is $v_{\rm imp}
\sim 0.7 v_{\rm bin}\,(a_{\rm bin}/0.07\;{\rm AU})^{-3/2}$. We have
checked that, for our range values of $a$ and $a_{\rm bin}$, the
inclusion of the above impact velocity has a mild impact on our
results (the maximum escapers' mean velocity is increased by $20\%$,
to $\sim 600$ km s$^{-1}$). A more detailed analysis, including for
instance the impact of an asymmetric supernova explosion, or the
possible change in the secondary's mass, is left for further
investigation.

The tidal disruption scenario remains a viable, if fine-tuned, unified
hypothesis for the S-stars, the runaway and the hypervelocity stars
\citep{peretshv09, lzy10a, lzy10b}. It requires compact binaries on
highly elliptical orbits prior to their tidal disruption. While
binaries forming and migrating in a gas disk may naturally become
compact (as shown in this study), they tend to have small or modest
eccentricities. Additional eccentricity excitation mechanisms, such as
resonant relaxation or a hypothetical intermediate-mass black hole,
need to be invoked. Similarly, the tidal disruption of disk binaries
gives stars with small inclination relative to the disk they formed
and migrated in. Again, additional mechanisms are required to account
for the apparent isotropic distribution of the S-stars. In both these
respects, the supernova disruption scenario requires less restricted
boundary conditions. It provides a natural way to form bound stars
with potentially large eccentricity and random inclination (depending
on the misalignment of the internal and orbital angular momentum
vectors of the binary prior to its disruption) from circular, compact
binaries. In the context of the hypervelocity stars, the supernova
disruption mechanism also provides a natural way for a $3-4 M_{\odot}$
star to be young at the point of ejection, whereas the tidal
disruption scenario does not select in favor of the young B stars.

%%%%%%%%%%%%
\acknowledgements
%%%%%%%%%%%%
We thank Alessia Gualandris, Javiera Guedes, Stephen Justham,
Katherine Kretke, Fr{\'e}d{\'e}ric Masset, Christopher Matzner, Sergei
Nayakshin, Hagai Perets, Enrico Ram{\'i}rez-Ruiz, Qingjuan Yu, and
Xiaojia Zhang for fruitful discussions and suggestions. The authors
are thankful to Stephen Justham, Fr{\'e}d{\'e}ric Masset and Hagai
Perets for their detailed comments on a first draft of this
manuscript. We also thank the anonymous referee for insightful
comments that helped improve the paper.  C.B. and J.C. are grateful to
the Kavli Institute for Astronomy and Astrophysics for its kind
hospitality and support during a portion of this work. Computations
were performed on the Pleiades and Laozi clusters at UC Santa
Cruz. This work is supported by NASA (NNX07AL13G, NNX07AI88G,
NNX08AL41G, NNX08AM84G), NSF (AST-0908807), the CAS Research
Fellowship for International Young Researchers, NSFC (10533030,
10821302), and the 973 Programme (2007CB815402).

\end{document}